\documentclass[12pt, oneside]{article}
\usepackage[left=1in,right=1in,top=1in,bottom=1in]{geometry}
\usepackage{graphicx}
\usepackage{amssymb}
\usepackage{amsmath}
\usepackage{amsthm}

\usepackage[frozencache,cachedir=.]{minted}
\usepackage{minted}
\usepackage{multirow}
\usepackage{pdflscape}
\usepackage{comment}

\usepackage{dcolumn} 
\newcolumntype{d}[1]{D{.}{.}{#1}}
\newcolumntype{e}[1]{D{,}{,}{#1}}

\usepackage{siunitx}

\usepackage{authblk}

\usepackage{color}
\usepackage[colorlinks,
            citecolor=blue,
            urlcolor=blue,
            linkcolor=blue]{hyperref}

\usepackage[utf8]{inputenc}

\usepackage[natbib=true,
            bibstyle=authoryear,
            citestyle=authoryear-comp,
            uniquename=false,
            uniquelist=false,
            maxcitenames=2,
            mincitenames=1,
            maxbibnames=7]{biblatex}
\renewbibmacro{in:}{\ifentrytype{article}{}{\printtext{\bibstring{in}\intitlepunct}}}

\usepackage{xpatch}
\xpatchbibmacro{name:andothers}{%
  \bibstring{andothers}%
}{%
  \bibstring[\emph]{andothers}%
}{}{}
\DeclareCiteCommand{\citeyear}
    {}
    {\bibhyperref{\printdate}}
    {\multicitedelim}
    {}

\DeclareCiteCommand{\citeyearpar}
    {}
    {\mkbibparens{\bibhyperref{\printdate}}}
    {\multicitedelim}
    {}

\title{Multivariate Modeling of Natural Gas Spot Trading Hubs Incorporating Futures Market Realized Volatility}
\author[1,2]{Michael Weylandt\footnote{To whom correspondence should be addressed: \href{mailto:michael.weylandt@rice.edu}{michael.weylandt@rice.edu}.}}
\author[1,2]{Yu Han}
\author[1,2]{Katherine B. Ensor}
\affil[1]{Department of Statistics, Rice University}
\affil[2]{Center for Computational Finance and Economic Systems, Rice University}

\usepackage{xspace}

\usepackage{bm}

\newcommand{\bI}{\bm{I}}

\usepackage{enumerate}
\usepackage{algorithm}
\usepackage{setspace}
\usepackage{rotating}

\RequirePackage[OT1]{fontenc}
\RequirePackage{amsthm,amsmath,amsfonts}
\usepackage{xspace}

\usepackage{booktabs}

\usepackage[parfill]{parskip}

\date{Last Updated: \today}

\usepackage{lineno}

\usepackage{subcaption}

\newcommand{\GARCH}{\texttt{GARCH}\xspace}

\newcommand{\RBGARCH}{\texttt{RBGARCH}\xspace}

\begin{document}
\maketitle
\singlespace
\begin{abstract}
Financial markets for Liquified Natural Gas (LNG) are an important and rapidly-growing segment of commodities markets. Like other commodities markets, there is an inherent spatial structure to LNG markets, with different price dynamics for different points of delivery hubs. Certain hubs support highly liquid markets, allowing efficient and robust price discovery, while others are highly illiquid, limiting the effectiveness of standard risk management techniques. We propose a joint modeling strategy, which uses high-frequency information from thickly-traded hubs to improve volatility estimation and risk management at thinly traded hubs. The resulting model has superior in- and out-of-sample predictive performance, particularly for several commonly used risk management metrics, demonstrating that joint modeling is indeed possible and useful. To improve estimation, a Bayesian estimation strategy is employed and data-driven weakly informative priors are suggested. Our model is robust to sparse data and can be effectively used in any market with similar irregular patterns of data availability.
\end{abstract}

 
\clearpage
\doublespace

\begin{refsection}
\section{Introduction}
Financial markets for Liquefied Natural Gas (LNG) are among the most important commodities markets in the United States (U.S.), with their importance rising as natural gas makes up an increasingly large share of U.S. energy consumption. Unlike other major commodities, the pricing of natural gas is largely driven by transportation and storage costs, which are reflected in the correlated price dynamics of different spot prices. In this paper, we develop a joint model for natural gas spot prices which takes advantage of the observable market structure and pools information across different time scales to more accurately forecast volatility for thinly traded spot prices. Our results  indicate that joint modeling of disparate natural gas spot prices is able to predict future volatility more accurately than standard univariate models.

Natural gas is a naturally-occurring mixture of hydrocarbons,
principally methane ($\text{CH}_4$), which is widely used for both large-scale commercial power generation and domestic use. In 2017, natural gas was used to generate approximately 1.273 petawatthours (PWh) or 1,273 billion killowatthours (kWh) of electricity in the U.S., comprising approximately 31.7\% of all electricity generated in the U.S. \citep{EIA-FAQ}. In 2015, natural gas surpassed coal as the principal source of electricity generation in the U.S. and is forecast to continue to make up a larger proportion of the U.S.'s electricity generation mixture in coming years \citep{AP:2015,EIA-STEO-2018-01}. This increase in usage has been primarily driven by the development of hydraulic fracturing (``fracking'') technology, which has driven down the cost of domestic production significantly by allowing natural gas to be extracted from shale efficiently and at relatively low cost \citep{BLS-2013}.

Natural gas is principally stored and transported in a cooled liquid form and, as such, so-called Liquefied Natural Gas markets are the main venue for large-scale commercial trade in natural gas. This trade is centered around a network of many standardized LNG transit and storage centers throughout the United States, colloquially known as ``hubs.'' (The use of the term ``hub'' may be somewhat confusing here since it does not imply that the storage center is centrally located or otherwise important, but we will use it throughout this paper as it is a standard terminology in these markets.) These hubs are connected by a nation-wide network of pipelines which connect the hubs to each other, as well as to important population centers and power generation plants, as shown in Figure \ref{fig:lng_pipeline_map} below. Like any commodity, LNG prices fluctuate unpredictably in response to market forces, sometimes rather significantly. Both upstream producers and downstream consumers, as well as investors in LNG markets from outside the supply chain, have a material interest in measuring and managing their exposure to these
fluctuations. 

Over-the-counter ``spot'' markets exist for almost every hub in the U.S., but the liquidity and transparency of these markets varies widely from hub to hub. For a few thickly-traded hubs, the spot market is highly liquid and supports a market of standardized futures contracts and other derivatives, not unlike equity (stock) or foreign exchange (FX) markets. The quality of these markets make them an attractive venue for speculators and market makers, who absorb risk from producers and consumers, improving overall market efficiency. For many other hubs, however, the spot market is thinly traded, with only a few large trades occurring each day, inhibiting price discovery, limiting the attractiveness of spot markets to third parties, and making risk management more difficult.

It has been observed by many authors that the use of high(er)-frequency ``realized'' volatility measures, based on intra-day price movements, significantly improves volatility estimation and allows for more efficient risk measurement, particularly in settings where a model based solely on daily data would otherwise be slow to react \citep{Anderson:2005,Hansen:2012,Hansen:2016}. It is therefore natural to ask whether similar techniques can be applied to LNG markets, particularly at thinly traded hubs where intra-day data may not be available. This paper answers that question in the affirmative, demonstrating that it is possible to use intra-day data from LNG futures markets to improve volatility estimation at thinly-traded hubs.

Taking the ``Realized Beta GARCH'' model of \citet{Hansen:2014} as our starting point, we develop a predictive single-factor multivariate volatility model for use in LNG markets. Our model is fully predictive and fully multivariate in nature, estimating volatility for multiple LNG spot and futures markets simultaneously. We adopt a Bayesian approach to parameter estimation in order to address two well-known difficulties of GARCH models: i) by incorporating information from a large pre-existing financial literature into our prior specifications, we are able to regularize our parameter estimation; and ii) by working in a fully Bayesian framework, we are able to coherently propagate uncertainty in our estimated parameters into our risk-management calculations.

The remainder of this paper is organized as follows: Section \ref{sec:mkts} provides additional background on LNG markets, after which Section \ref{sec:lit} reviews related work. Section \ref{sec:data} describes the data set used in the following analysis and highlights several stylized facts that motivate our proposed model. Section \ref{sec:model} specifies our proposed model in detail, with Section \ref{sec:priors} providing additional detail on suggested priors.  Section \ref{sec:application} demonstrates the usefulness of our model with an application to U.S. LNG markets, highlighting the improved accuracy on standard risk-management benchmarks. Finally, Section \ref{sec:disc} closes with a discussion and a consideration of some directions for future work. Supplemental materials give additional background, data description, and results.

\subsection{U.S. Natural Gas Markets} \label{sec:mkts} 
\begin{figure}
  \centering
  \includegraphics[width=4.5in,height=2.6in]{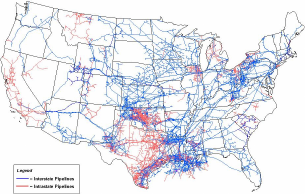}
  \caption{U.S.~LNG Pipelines (as of 2008). There are approximately 3 million miles of LNG pipelines in the United States, connecting LNG storage facilities with customers. The majority of these pipelines concentrated along the Gulf of Mexico and in western Pennsylvania / northern West Virginia. Note the large number of interstate pipelines originating at or near Henry Hub in southern Louisiana. Figure originally published by the \citet{EIA-MAP}.}
  \label{fig:lng_pipeline_map}
\end{figure}

Conceptually, the simplest mechanism for trading LNG is through so-called ``spot'' markets. Trades in this market lead to (essentially) immediate exchange of cash for delivery of LNG to the purchaser's account at a pre-determined hub. Historically, the most important of these hubs is the so-called ``Henry Hub,'' located immediately outside the town of Erath in southern Louisiana. While LNG markets have expanded nationwide, Henry Hub, and the associated distribution network, remains a key transit nexus and price reference. The transit and storage facilities at Henry Hub are connected to several major interstate LNG pipelines, facilitating easy transit throughout the U.S., as shown in Figure \ref{fig:lng_pipeline_map}. As such, spot prices at Henry Hub are generally understood to serve as a proxy for U.S. LNG market prices more broadly. 

As with many commodities, markets for LNG are heavily financialized, with a wide array of futures, options, and other derivative securities being heavily traded. Because of the central role Henry Hub plays in spot markets, the vast majority of these derivatives are directly or indirectly based on Henry Hub spot prices. The New York Mercantile Exchange introduced futures contracts based on Henry Hub prices in 1990; these contracts are traded at a variety of maturities up to 18 months at an average daily volume of approximately \$14 billion USD and comprise the third largest commodity market in the U.S., exceeded only by crude oil and gold \citep{CME-2018-Q1}. In recent years, the Intercontinental Exchange has also become a leading venue for LNG derivatives, currently serving as the venue for approximately 40\% of total open interest in Henry Hub-referenced futures \citep{ICE-LNG}. Non-U.S. hubs are of increasing interest, recognizing the evolution and growth of international LNG markets \citep{EIA-AsiaHubs:2017}, but we restrict our focus to hubs in the continental United States.

Unlike equity or bond markets, the mechanisms by which LNG is stored and transported have a significant effect on price behaviors. \citet{Mohammadi:2011} gives a detailed survey of the pricing structure from well-heads (extraction) to end-consumers.  The relationship between LNG spot and futures prices was investigated by \citet{Ghoddusi:2016} who found, \emph{inter alia} that short-maturity futures are Granger-causal for physical prices, a fact which is consistent with our findings on the usefulness of futures realized volatility in spot price volatility modeling. To the best of our knowledge, the subject of this paper -- the relationship between the instantaneous volatilities of different spot prices -- has not been previously examined in the literature.

\subsection{Previous Work} \label{sec:lit}

Financial time series, including LNG prices, exhibit complex and well-studied patterns, notably significantly non-constant patterns of volatility (heteroscedasticity). This motivates the use of instantaneous volatility, typically understood as the instantaneous standard deviation of returns, as an explicit and important quantity in any realistic model. Direct measurement of volatility, however, is a difficult task, as volatility can change as rapidly or more rapidly than prices can be observed. To address this limitation, two parallel lines of research have been developed in the econometrics literature. The first, latent volatility process models, posit dynamics for the unobserved volatility process and attempt to estimate the parameters of those dynamics directly from returns data; these models are typically further sub-divided into stochastic volatility \citep{Kim:1998,Vankov:2019} and (generalized) autoregressive conditional heteroscedasticity (GARCH), depending on the form of the assumed dynamics \citep{Engle:1982,Bollerslev:1986}. We focus on GARCH models here as they form the basis of our proposed approach. These models typically assume a daily volatility $\sigma_t^2$ which is never observed directly, but can be estimated over long periods of time using many realizations of the squared daily return $r_t^2 =(\log P_t / P_{t-1})^2$, essentially using only the estimator $\hat{\sigma_t}^2 \approx (r_t - \overline{r})^2$ and using a model to pool estimates through time.  Given the fact that these models allow the instantaneous volatility to change rapidly and have so little information per time point, they are often more effective for recovering the parameters of the dynamics than for estimating the instantaneous volatility at a given time point. Despite this, these models have proven immensely popular in the econometric literature, spawning an enormous number of variants and refinements; see, \emph{e.g.}, the thirty-page glossary of \citet{Bollerslev:2010} or the review of \citet{Bauwens:2006}. 

The second line of research, so-called ``realized volatility'' models, attempts to enlarge the information set used to estimate the integrated (average) volatility over a fixed period of time. Early work in this direction incorporated additional commonly-cited summary statistics, such as the open (first) price, the intra-day trading range, or combinations thereof \citep{Rogers:1991}. \citet{Yang:2000} derived an optimal (minimum-variance unbiased) estimator allowing non-zero drift and overnight jumps,
which can be interpreted as a weighted sum of the overnight, intra-day, and Rogers-Sachell \citeyearpar{Rogers:1991} volatility estimators. When appropriately tuned, the Yang-Zhang estimator is up to fourteen times more efficient than the standard close-to-close volatility estimate \citep{Yang:2000}, though its assumption of continuous price dynamics leads it to slightly underestimate volatility in the presence of jumps. Moving beyond ``OHLC'' (Open, High, Low, Close) data, more recent work uses even higher-frequency data, correcting for the idiosyncrasies of market micro-structure, but these adjustments are typically developed for equities and are quite sensitive to market microstructure, so we do not pursue them here.

Given the successes of both of these lines of research, it is natural to combine them into a single model, in which instantaneous volatility is still assumed to evolve according to some form of GARCH dynamics, but for which we have more accurate estimators than the squared daily return. The Realized GARCH framework introduced by P.R.~Hansen and co-authors \citep{Hansen:2012,Hansen:2016} is a recent entry in this vein, which allows for slightly more general specifications than the previously proposed Multiplicative Error Model (MEM) or High-Frequency Based Volatility (HEAVY)  frameworks \citep{Engle:2006,Shephard:2010}. In particular, we build upon the ``realized beta GARCH'' model of \citet{Hansen:2014} which proposes a model with two series, termed the ``market'' and the ``asset,'' where the market has a realized volatility measure available but the asset does not, and uses information from the market to improve estimation for the asset. We adapt this approach for predictive risk management in the LNG market, describing a coherent Bayesian estimation strategy complete with data-driven priors and an out-of-sample VaR forecasting scheme. \citet{Contino:2017} previously considered Bayesian approaches to estimating realized GARCH models, though they do not consider the multivariate realized beta GARCH model that we use here.

The Network GARCH model of \citet{Zhou:2018} is similar to the model we present in Section \ref{sec:model}, but they assume that the observations are conditionally independent and have the squared return of one asset directly influence the latent volatility of another asset. As we show below, daily returns of LNG spot prices are highly correlated, so the Network GARCH is clearly misspecified for our application. Interpreted as a network, our model encodes the marginal volatilities as a star graph, with Henry Hub at its center and other nodes connected only to it. We leave the problem of allowing and estimating a more general graph structure to future work.

\section{Empirical Characteristics of LNG Prices} \label{sec:data}

Commercial data vendors recognize over one-hundred and fifty tradeable spot LNG prices in the continental U.S. and Canada (see, \emph{e.g.}, the \texttt{BGAS} screen on a Bloomberg terminal). Not surprisingly, a full data history is not available for each of these price series. We curate a data set of forty spot prices for which we were able to obtain a full daily price history over the ten-year period ranging from January 2006 to December 2015 (2209 trading days). This period roughly aligns with the third subsample period identified by \citet{Li:2019}, who describes it as a large, active, and growing market with lower levels of volatility than observed during the initial emergence of LNG markets. Additionally, we use OHLC data from Henry Hub futures at one, two, three, six, nine, and twelve month maturities to calculate the market return and realized volatility measures. 
We highlight several key properties of our data set below and give additional details in Section \ref{app:descrip} of the Supplementary Materials. Section \ref{app:series} of the Supplementary Materials describes our data curation and gives replication instructions.

\begin{figure}[htb]
  \centering
  \includegraphics[width=6.5in,height=2in]{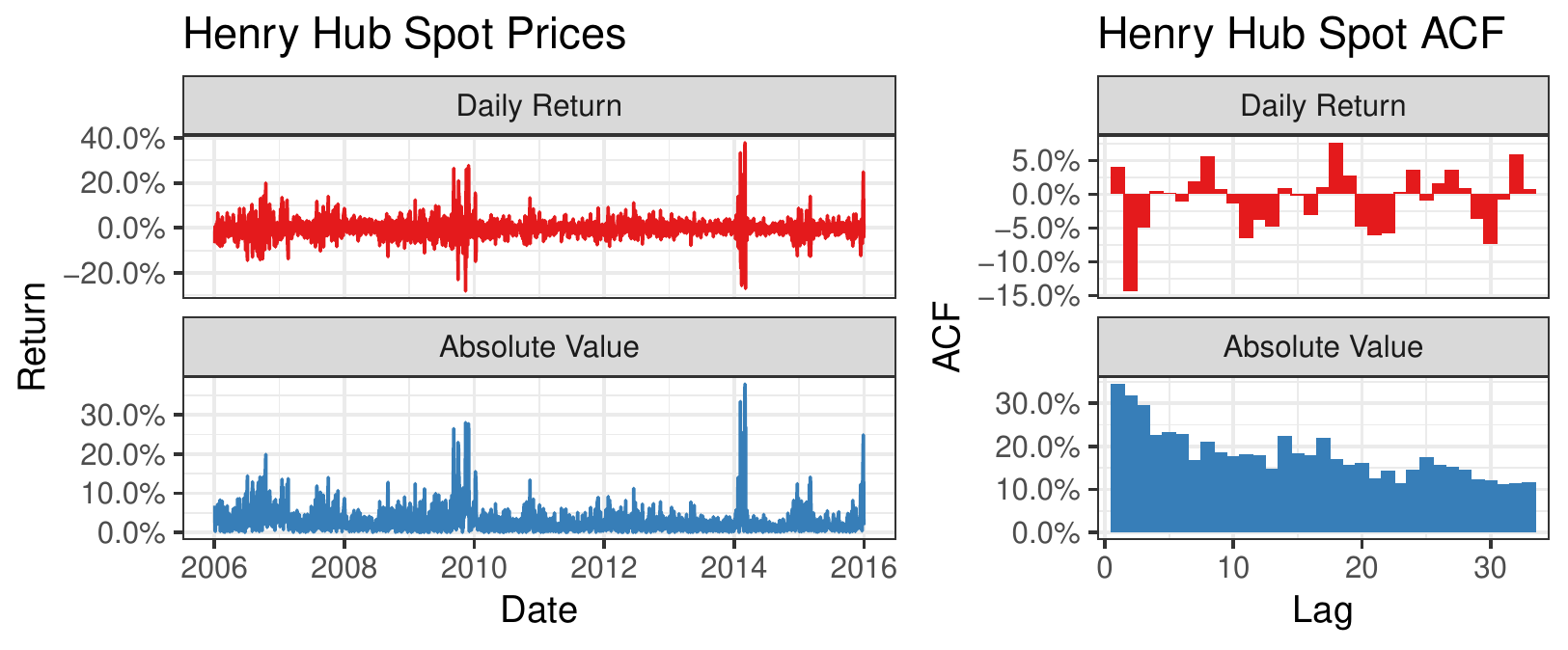}
  \caption{Henry Hub spot returns. Henry Hub clearly exhibits many of the same properties as equity returns: volatility clustering, heavy-tails, and autocorrelation in the second moment.}
  \label{fig:hhub_facts}
\end{figure}

LNG markets exhibit many of the same well-known characteristics as equity markets, including asymmetric heavy-tailed returns, volatility clustering, and long-lasting autocorrelation of squared returns. Figure \ref{fig:hhub_facts} shows the daily return of the Henry Hub spot, as well as its absolute value and autocorrelation function. From this, it is clear that Henry Hub exhibits volatility clustering and log-lasting autocorrelation in its second moment. This series exhibits strong evidence of heavy tails (sample excess kurtosis of 11.3), but only limited evidence of skewness (sample skewness 0.72). Other spot prices exhibit similar behaviors. 

\begin{figure}[htb]
  \centering
  \includegraphics[width=6.5in,height=2.2in]{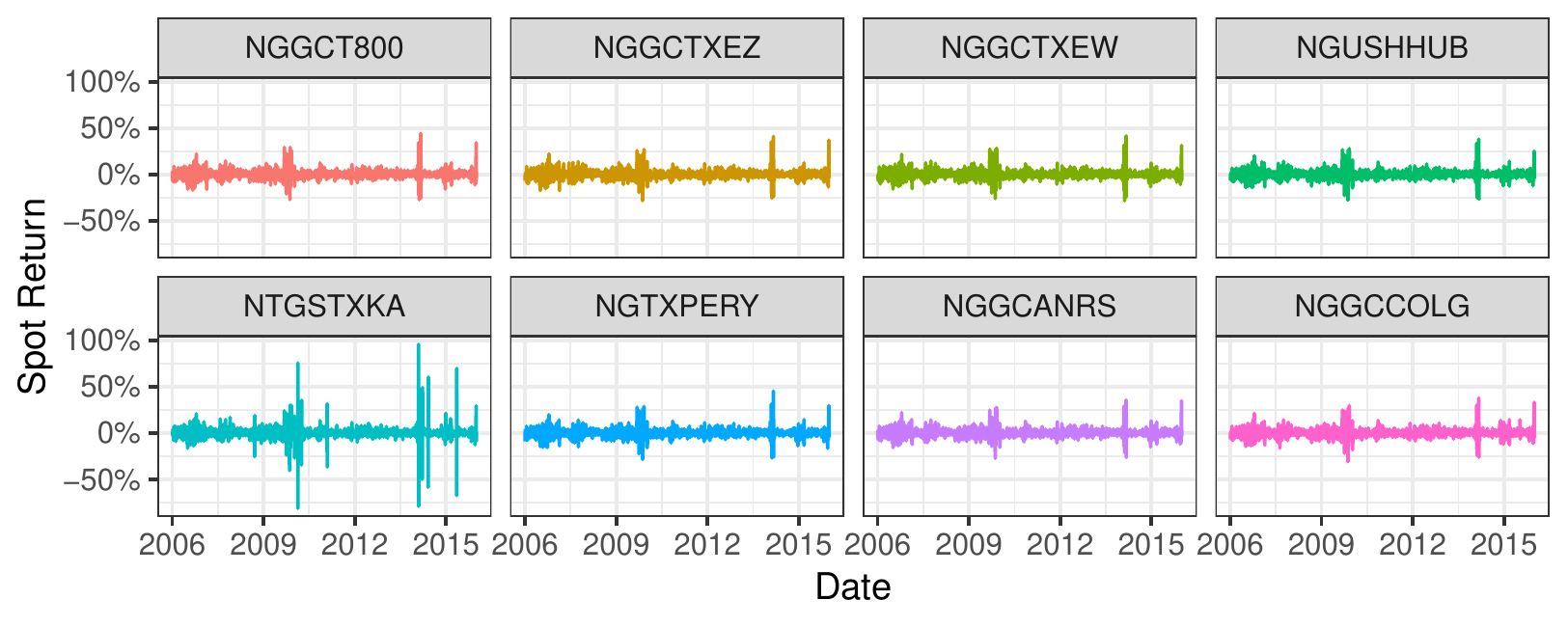}
  \caption{Return series of several LNG spots. (Details of each spot series are given in Section \ref{app:series} of the Supplementary Materials.) Returns are clearly highly correlated across series, with the major volatility events (late 2009 and early 2014) being observed at all hubs. This high degree of correlation suggests that a single-factor model will suffice.}
  \label{fig:spots_together}
\end{figure}

Spot prices at different trading hubs tend to move together, with shocks at one hub being quickly reflected in connected hubs and eventually throughout the entire LNG market. As can be seen in Figure \ref{fig:spots_together}, the daily price returns at different spots are clearly highly correlated and exhibit similar volatility clustering patterns. For the 40 spot prices considered in Section \ref{sec:application}, the first principal component explains approximately 74\% of the total Spearman covariance, suggesting that a single market factor drives most of the observed variability.

\begin{figure}[htb]
  \centering
  \includegraphics[width=6.5in,height=3.2in]{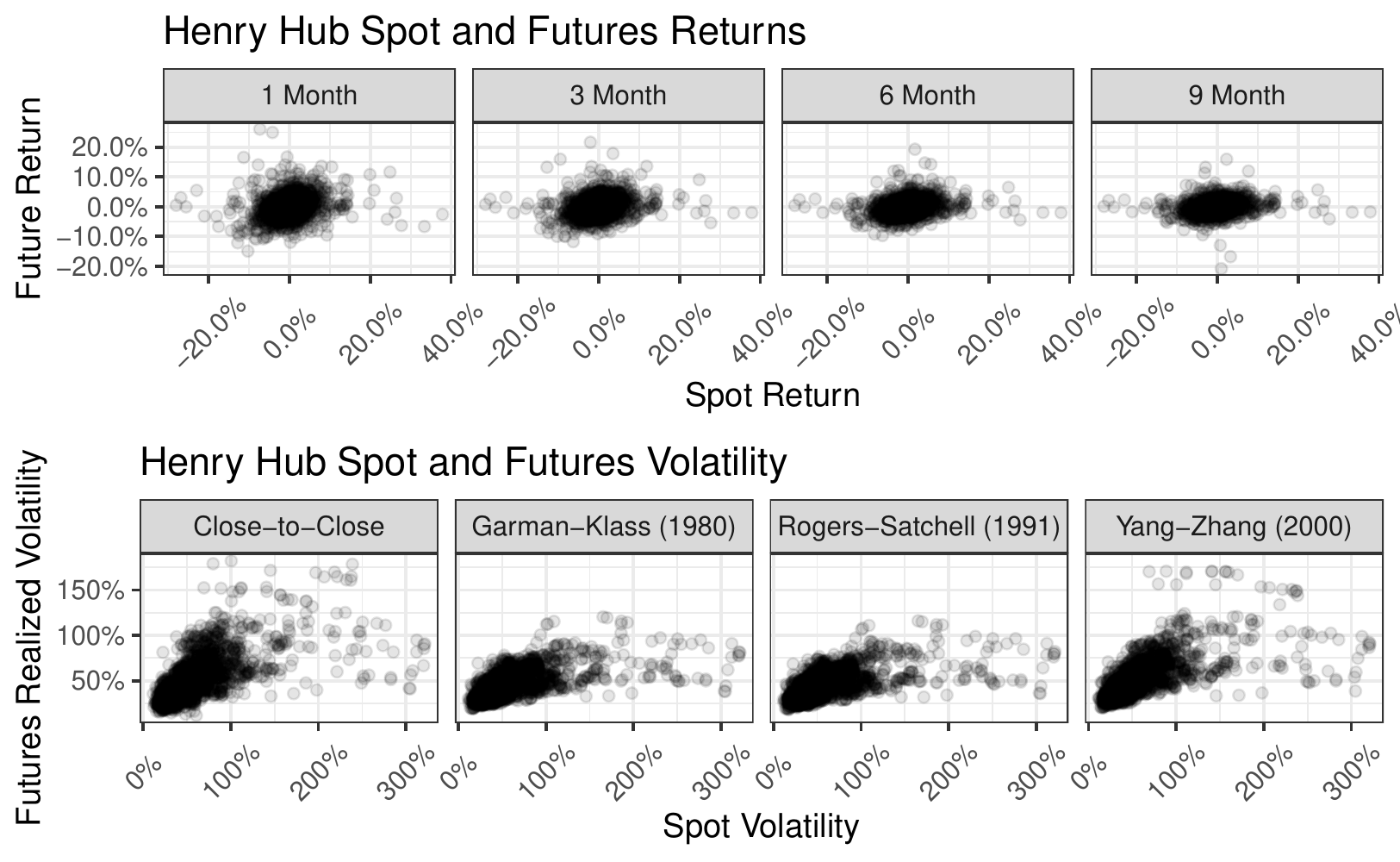}
  \caption{Comparison of Henry Hub Spot and Futures. As shown in the top panel, spot and futures returns are moderately correlated (25-29\%), particularly at shorter maturities (29.2\%). As shown in the bottom panel, spot and futures volatitlities are highly correlated (74-99\%) for all of the realized volatility measures considered, including the close-to-close, Garman-Klass \citeyearpar{Garman:1980}, Rogers-Satchell \citeyearpar{Rogers:1991}, and Yang-Zhang \citeyearpar{Yang:2000} estimators. In both cases, we observe strong correlation between the low frequency spot data and the high frequency futures data.}
  \label{fig:futures}
\end{figure}

LNG spot prices, particularly those for Henry Hub, are also closely connected to the associated futures market. As can be seen in the top panel of Figure \ref{fig:futures}, Henry Hub spot and futures are highly correlated, with the correlation highest for the shortest maturity futures. Similarly, spot and futures volatility estimates are also highly correlated, as shown in the bottom panel of Figure \ref{fig:futures}. Because the futures market is typically more liquid than the spot market, we can use the additional information provided by this market to improve estimation of spot volatilities. 

Putting these results together, we see that the spot LNG market exhibits equity-like volatility patterns with strong correlations among different spots and between the spot and futures markets. Spot prices exhibit quite heavy tails, but relatively little skewness, suggesting the use of a skewed observation distribution rather than an asymmetric GARCH formulation. Finally, the dominant leading principal component, indicative of market-wide volatility shifts, suggests that a single factor model will perform well. We will incorporate each of these observations as we develop our proposed model in the following section. 

\section{Model Specification} \label{sec:model}
In this section, we introduce our model piecewise, giving motivation for our modeling choices as we proceed. The complete specification is given in Equation \eqref{eqn:model} at the end of this section. Conceptually, our model has three major components: a standard GARCH specification for the volatility at Henry Hub ($\sigma_{M, t}$), multiple realized volatility measures ($\varsigma_{j, t}$), and an augmented GARCH specification for the non-Henry Hub spots which includes an additional term ($\sigma_{M, t}^2$) to capture market-wide changes in the volatility. Because our eventual aim is risk management, our model is fully predictive: in particular, volatilities at time $t$ ($\sigma_{M, t}, \sigma_{i, t}$) depend only on past values, allowing for out-of-sample forecasting, rather than having $\sigma_{M, t}$ influence $\sigma_{i, t}$ as in the specification of \citet{Hansen:2014}.

We adopt a slight variant of a standard linear GARCH(1, 1) specification for the instantaneous volatility at Henry Hub: 
\[\sigma_{M, t}^2 = \omega_M + \gamma_M \sigma_{M, t-1}^2 + \sum_{j=1}^J\zeta_j \varsigma_{j, t - 1}^2 + \tau_{M, 1} |r_{M, t - 1}| + \tau_{M, 2}(r_{M, t-1} - \mu_M)^2\]
where $r_{M, t}$ is the daily Henry Hub spot price return at $t$, $\mu_M$ is the long-run average return, and $\varsigma_{j, t - 1}$ is the value of the $j$\textsuperscript{th} realized volatility measure at time $t - 1$. \citet{Chan:2016} note that, contrary to crude oil markets, the presence of a leverage term has minimal predictive value in LNG markets. As such, we follow \citet{Hansen:2012} and use a second-order Hermite polynomial in $|r_{\cdot, t}|$ to allow for flexible modeling of leverage, rather than enforcing leverage directly in our specification. Note that, since we are using a linear rather than log-linear formulation, we include an absolute value term to disallow negative volatilities. The realized volatility term ($\zeta_j^2$) allows our model to indirectly incorporate aspects of volatility captured by the realized volatility measurements which are not otherwise captured by the close-to-close return.

The individual asset volatilities follow a similar specification, though we introduce an additional ``coupling'' term to capture market-wide changes in volatility where Henry Hub is used as a proxy for the market as a whole:
\[\sigma_{i, t}^2 = \omega_i + \gamma_{i} \sigma_{i, t-1}^2 + \tau_{i, 1} |r_{i, t-1}| + \tau_{M, 2}(r_{i, t-1} - \mu_i)^2 + \beta_{i} \sigma^2_{M, t-1} \quad \forall i=1, \dots, I\] where $r_{i, t}$ is the return of (non-Henry Hub) spot price $i$ at time $t$. The coupling parameter $\beta_i$ measures the influence of Henry Hub volatility on the other hubs. If $\beta_i = 0$, then Henry Hub volatility is minimally informative of the volatility at hub $i$ (conditionally independent given the past observables), while a large value of $\beta_i$ indicates that the secondary hub experiences significant ``spill-over'' volatility from Henry Hub. 

Conditional on these volatilities, we assume that the daily returns follow a multivariate skew-normal distribution, though any skewed elliptical distribution, \emph{e.g.,} the multivariate skewed $t$ distribution, could be used as well. In other words,
\begin{align*}
\vec{r}_t &= \begin{pmatrix} r_{M, t} \\ \{r_i, t\}_{i=1}^I \end{pmatrix}\sim \text{Multi-Skew-Normal}\left(\alpha, \mu, \Sigma_t \right) \\ 
\Sigma_t &= \text{diag}\left(\sigma_{M, t}, \{\sigma_{i, t}\}_{i=1}^k\right)\; \Omega\; \text{diag}\left(\sigma_{M, t}, \{\sigma_{i, t}\}_{i=1}^k\right)
\end{align*}
where $\alpha$ and $\mu$ are fixed (non-time varying) skewness and mean parameters and $\Omega$ is the fixed (non-time varying) correlation of returns. Because our data exhibits relatively low degrees of skewness, we do not impose strong skewness through our model specification, instead allowing for the possibility of skewness by combining the linear GARCH specification with a skew-normal return distribution \citep{Azzalini:1999}. Empirically, we have found this scheme to perform better than the log-linear specification described by \citet{Hansen:2012}, but, as they note, the success of the realized GARCH framework does not depend critically on the specification used. 

Finally, we include an explicit model for the realized volatility measures: 
\[\varsigma_{j, t} \sim \mathcal{N}(\xi_j + \phi_j \sigma_{M, t} + \delta_{j, 1} |r_{M, t}| + \delta_{j, 2} r_{M, t}^2, \nu_j^2) \quad \forall j=1, \dots, J.\]
As before, we incorporate a leverage effect using a second order Hermite polynomial. We emphasize that the realized volatility measures need not be unbiased estimates of the true market volatility at time $t$. The bias ($\xi_j$) and scaling ($\phi_j$) factors allow for mis-scaled or mis-aligned volatility measures. In other words, an intra-day volatility measure does not need to be re-scaled to daily volatility. This is especially important in our application below, where realized volatility measures from futures markets are used for spot markets. (See Figure \ref{fig:futures} for evidence of different scalings.) The bias and scaling terms allow these estimates to inform our estimation, even though our model does not directly account for the particular dynamics of the futures markets (\emph{i.e.}, cost of carry, interest rate dynamics, \emph{etc.}; see the discussion by \citet{Li:2019}). The use of drift-adaptive volatility measure of \citet{Yang:2000} also helps to ameliorate these effects.). As we will see, our model is able to adapt to the mismatch between the futures and spot markets, demonstrating that, despite this discrepancy, the futures market realized volatility does indeed constitute a useful addition to spot models.

Putting these pieces together, we obtain the complete specification: 
\begin{align}
\vec{r}_t &= \begin{pmatrix} r_{M, t} \\ \{r_i, t\}_{i=1}^I \end{pmatrix}\sim \text{Multi-Skew-Normal}\left(\alpha, \mu, \Sigma_t \right) \nonumber \\
\Sigma_t &= \text{diag}\left(\sigma_{M, t}, \{\sigma_{i, t}\}_{i=1}^k\right)\; \Omega\; \text{diag}\left(\sigma_{M, t}, \{\sigma_{i, t}\}_{i=1}^k\right) \nonumber \\
\sigma_{M, t}^2 &= \omega_M + \gamma_{M} \sigma_{M, t-1}^2 + \sum_{j=1}^J \zeta_j \varsigma_{j, t-1}^2 + \tau_{M, 1} |r_{M, t-1}| + \tau_{M, 2}(r_{M, t-1} - \mu_M)^2 \label{eqn:model}\\
\sigma_{i, t}^2 &= \omega_i + \gamma_{i} \sigma_{i, t-1}^2 + \tau_{i, 1} |r_{i, t-1}| + \tau_{i, 2}(r_{i, t-1} - \mu_i)^2 + \beta_{i} \sigma^2_{M, t-1} \quad \forall i=1, \dots, I \nonumber \\
\varsigma_{j, t} &\sim \mathcal{N}(\xi_j + \phi_j \sigma_{M, t} + \delta_{j, 1} |r_{M, t}| + \delta_{j, 2} r_{M, t}^2, \nu_j^2) \quad \forall j=1, \dots, J \nonumber
\end{align}
where
\begin{description}
  \item[$r_{M, t}$] is the market return at time $t$ (observed);
  \item[$r_{i, t}$] is the return of (non-Henry Hub) spot price $i$ at time $t$ (observed);
  \item[$\alpha, \mu$] are fixed (non-time-varying) return parameters (unobserved);
  \item[$\Sigma_{t}$] is the conditional variance at time $t$ (unobserved);
  \item[$\sigma_{M, t}$] is the instantaneous market volatility at time $t$ (unobserved);
  \item[$\sigma_{i, t}$] is the instantaneous volatility of spot price $i$ at time $t$ (unobserved);
  \item[$\beta_i$] measures the effect of market volatility on the volatility of spot price $i$ (unobserved);
  \item[$\zeta_i$] measures the influence of realized volatility measure $j$ on market volatility (unobserved);
  \item[$\omega, \gamma, \tau$] are fixed (non-time-varying) GARCH parameters (unobserved);
  \item[$\varsigma_{j, t}$] is a realized measure of market volatility at time $t$ (observed); and
  \item[$\xi, \phi, \delta, \nu$] are fixed (non-time-varying) parameters of the realized volatility measurement (unobserved).
\end{description}

\subsection{Bayesian Estimation and Prior Selection} \label{sec:priors}
To estimate the parameters of our model, we adopt a Bayesian approach. As we will see below, this poses several advantages, perhaps the most obvious of which is that it allows us to incorporate information from the large GARCH literature in the form of a prior. A less obvious, but equally important, advantage is that the Bayesian framework allows for coherent propagation of parameter uncertainty into subsequent analyses. In the financial context, this typically produces more extreme, but ultimately more accurate, risk measures \citep{Ardia:2017}. 


The choice of priors is fundamental to any Bayesian model.  Rather than developing priors based on theoretical considerations, we derive priors from a related but independent data set where possible. In particular, we fit a (univariate) realized GARCH model to the returns of the S\&P 500, an index of major U.S. stocks, and use the maximum likelihood estimates to center our priors for LNG. This process is repeated over 250 of randomly-chosen year-long periods during our sample period (2006-2015) and the median estimate is used as the mean of the prior. The prior standard deviation is matched to ten times the median absolute deviation of the MLEs. This yields the prior specifications shown in Table \ref{tab:priors}. For priors which can be well estimated from the data, \emph{e.g.}, the mean return $\mu$, or the fixed correlation matrix, $\Omega$, we use a standard weakly informative prior.
Throughout, we also constrain our priors, and hence our posterior, to ensure stationarity of the underlying process.

\begin{table}[htbp]
\centering
\scalebox{0.8}{
\begin{tabular}{cccc}
\toprule
{\bf Parameter(s)} & {\bf Interpretation}  & {\bf Calibration Strategy} & {\bf Prior} \\
\midrule
$\mu$ & Mean Return & \multirow{4}{*}{Weakly Informative} & $\mathcal{N}(0, \bI)$ \\
$\beta$ & Volatility coupling &  & $\mathcal{N}(0, \bI)$ \\
$\Omega$ & Return Correlation &  & Haar Measure (Uniform) \\
$\alpha$ & Return Skewness &  & $\mathcal{N}(0, \bI)$ \\
\midrule
$\omega$ & Long-Run Volatility Baseline & \multirow{9}{*}{SPY MLE Calibration} & $\mathcal{N}(0.002, 0.025^2)$ \\
$\gamma$ & Volatility Persistence &  & $\mathcal{N}(0.8, 0.6^2)$ \\
$\tau_1$ & First-Order GARCH Effect & & $\mathcal{N}(0, 0.01^2)$  \\ 
$\tau_2$ & Second-Order GARCH Effect & & $\mathcal{N}(0.1, 0.7^2)$  \\ 
$\xi$ & Realized Volatility Bias & & $\mathcal{N}(0.02, 0.6^2)$ \\ 
$\phi$ & Scaling of Realized Volatility  & & $\mathcal{N}(15, 60^2)$ \\ 
$\nu$ & Noise of Realized Volatility (R.V.) & & $\mathcal{N}(0.05, 0.25^2)$ \\ 
$\delta_1$ & R.V.~First-Order Leverage Effect  & & $\mathcal{N}(1.15, 8^2)$  \\
$\delta_2$ & R.V.~Second-Order Leverage Effect  & & $\mathcal{N}(1.15, 14^2)$  \\
\bottomrule
\end{tabular}}
\caption{Prior Specifications. Parameters for first-order behavior (mean and correlation of returns) are set using standard weakly informative priors. Priors for GARCH dynamics are calibrated to S\&P500 dynamics over the sample period (2006-2015). The ``coupling parameters'' $\beta$ which are the primary focus of this paper are given independent weakly informative $\mathcal{N}(0, 1)$ priors.}
\label{tab:priors}
\end{table}

To ensure the reasonableness of our priors, we generate simulated trajectories with GARCH parameters drawn from our priors as a ``prior predictive check.'' On visual inspection of simulated price series, we observe that our priors imply realistic GARCH dynamics, though the volatility is higher than what we actually observe, consistent with our use of weakly informative (wide) priors on the GARCH parameters, which allow for higher-than-realistic volatilities. 

Table \ref{tab:prior_pc} compares the 90\% prior predictive intervals of several relevant statistics against the S\&P 500 data used to calibrate our priors and against Henry Hub spot returns. Our priors imply a kurtosis below the observed kurtosis of the S\&P 500 and Henry Hub. This divergence appears to be driven by the model not accounting for large exogenous jumps rather than an incorrect GARCH specification. Overall, we see that our priors are realistic, even though they often imply a higher ``volatility of volatility'' than actually observed.

\begin{table}[htbp]
\centering
\scalebox{0.8}{
\begin{tabular}{ccS[table-format=2.2]S[table-format=2.2]}
\toprule
{\bf Test Statistic} & {\bf 90\% Interval} & {\bf S\&P 500} & {\bf Henry Hub Spot} \\ 
\midrule
Mean Return & [-0.025, 0.022] & 0.00 & -0.00 \\ 
Mean Absolute Return & [0.058, 0.519] & 0.01 & 0.03 \\ 
Standard Deviation of Returns & [0.073, 0.761] & 0.01 & 0.04 \\ 
Skewness of Returns & [-0.545, 0.484] & -0.15 & 0.72 \\ 
Kurtosis of Returns & [2.839, 13.362] & 17.78 & 14.38 \\ 
 Fraction of Positive Returns & [0.466, 0.535] & 0.55 & 0.47 \\ 
One-Day Autocorrelation of Returns & [-0.108, 0.112] & -0.08 & 0.04 \\ 
Two-Day Autocorrelation of Returns & [-0.092, 0.089] & -0.08 & -0.14 \\ 
One-Day Partial Autocorrelation of Returns & [-0.094, 0.08] & -0.09 & -0.15 \\ 
One-Day Autocorrelation of Squared Returns & [0.014, 0.56] & 0.18 & 0.27 \\ 
Two-Day Autocorrelation of Squared Returns & [-0.005, 0.433] & 0.45 & 0.29 \\ 
One-Day Partial Autocorrelation of Squared Returns & [-0.060, 0.239] & 0.43 & 0.23 \\ 
\bottomrule
\end{tabular}}
\caption{Prior Predictive Checks for Model \eqref{eqn:model}: 90\% Prior Predictive Intervals and observed S\&P 500 and Henry Hub values. The 90\% Prior Predictive Intervals cover the observed values for each of the standard summary statistics except those based on unconditional volatility (standard deviation and kurtosis of returns), where the priors imply a slightly higher volatility than actually observed. As we will see in Section \ref{sec:application}, this is consistent with the out-of-sample evaluation of our model, which is generally accurate, but slightly conservative.}
\label{tab:prior_pc}
\end{table}

\subsection{Computational Concerns}
Full Bayesian inference for Model \eqref{eqn:model} requires use of a Markov Chain Monte Carlo (MCMC) sampler. Because the latent volatilities, $\sigma_{M, T}$ and $\sigma_{i, T}$, are typically highly correlated across time, Hamiltonian Monte Carlo samplers are particularly well suited for this problem. We make use of the \texttt{Stan} probabilistic programming language and its variant of the No-U-Turn Sampler \citep{Carpenter:2017,Hoffman:2014}. While \texttt{Stan} is able to sample the posterior relatively efficiently for this problem, simultaneous inference for a large number of assets, \emph{e.g.}, the 40 spot prices considered in the next section, is still computationally quite demanding. To avoid this, we take advantage of the factorizable structure of Model \eqref{eqn:model} and compute partial posteriors for each asset separately and combine our partial posteriors to obtain an approximate posterior. The parameters of the joint return distribution can then be estimated efficiently after by treating the returns as independent samples after standardization by the estimated instantaneous volatilities. Additional details of our Markov Chain Monte Carlo procedure and convergence diagnostics are included in Section \ref{app:computation} of the Supplementary Materials.

\section{Application to Financial Risk Management} \label{sec:application}
We demonstrate the effectiveness of our model with an application to LNG spot prices at non-Henry hubs. We fit our model, under the priors described in Section \ref{sec:priors}, on a 10 year historical period, refitting every 50 days using a 250-day look-back window. This periodic refitting strategy implicitly allows for time-varying coefficients, unlike the approach of \citet{Hansen:2014}, who only allow for time-varying correlations ($\Omega$) with a parametric specification for the temporal dynamics. Unless otherwise stated, all volatility measures are posterior expectations taken over 4000 posterior samples. Throughout, we compare the results of Model \eqref{eqn:model}, which we denote as \RBGARCH for realized beta GARCH, with a pair of standard skew-normal GARCH(1, 1) models, which we denote as \GARCH. To ensure a fair comparison, the same model specification and priors are used for both \GARCH and \RBGARCH. While the \RBGARCH model is able to convincingly outperform the \GARCH model in-sample, the superior out-of-sample risk-management performance is particularly compelling evidence as to the usefulness of the realized beta GARCH model in LNG markets. Additional comparisons are given in Section \ref{app:additional_results} of the Supplementary Materials.

\subsection{In-Sample Model Fit}
We first compare the in-sample model fit of the \GARCH and \RBGARCH models. To assess improvement in in-sample performance, we use \citeauthor{Watanabe:2010}'s \citeyearpar{Watanabe:2010} \emph{Widely Applicable Information Criterion} (WAIC) . Essentially a Bayesian analogue of AIC, WAIC is an approximate measure of the leave-one-observation-out predictive performance of a model. The likelihood of the \RBGARCH model contains terms for both the daily returns and the realized volatility measure. To ensure a fair comparison with the \GARCH model, we calculate the WAIC of the \RBGARCH model on a \emph{partial} likelihood, using only the likelihood of the daily returns. The partial likelihood is analogous to the complete likelihood of the \GARCH model, which does not include a term for the realized volatility measure, and allows for an apples-to-apples WAIC comparison of the two models. 

\begin{figure}[ht]
  \centering
  \includegraphics[width=\textwidth]{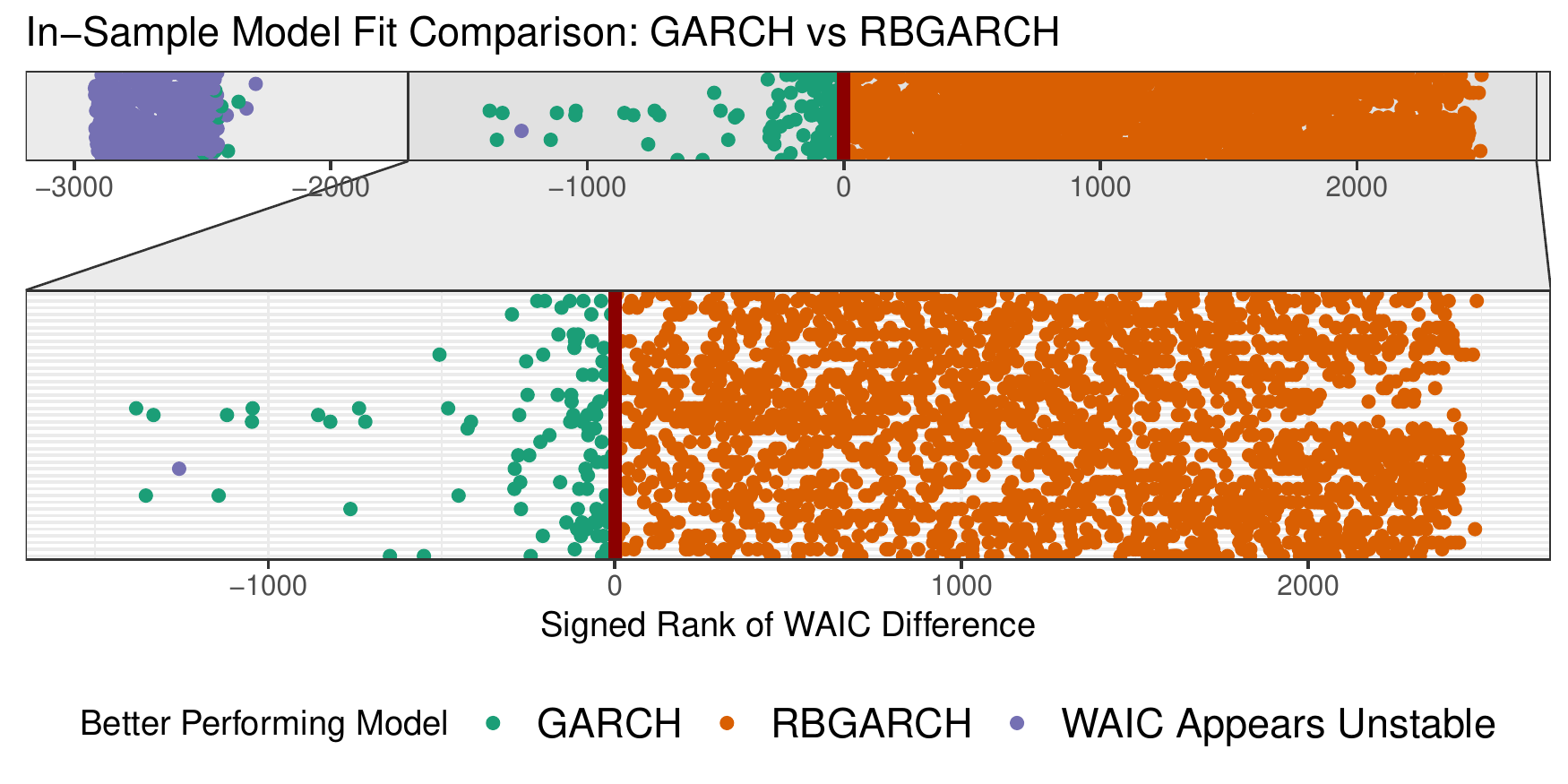}
  \caption{Signed-rank-transformed differences of 40 spots over 73 periods. Ignoring periods where our WAIC estimates are corrupted by outliers (purple), the \RBGARCH model performs better the \GARCH model in 2333 out of 2444 (95.4\%) sub-samples.}
  \label{fig:waic}
\end{figure}

In Figure \ref{fig:waic}, we compare the in-sample performance of the \GARCH and \RBGARCH models using a signed-rank transform to account for the highly non-normal sampling distribution of the WAIC differences. The left portion of Figure \ref{fig:waic} shows a small number of sub-periods where the WAIC comparison appears to strongly favor the \GARCH model. A closer investigation of these sub-periods reveals that these irregularities are caused by short-lived outliers in the underlying data which corrupt the WAIC estimates. \citet{Vehtari:2017} suggest that if the posterior variance of posterior log-predictive density exceeds 0.4 for any single observation, WAIC does not provide a robust estimate to out-of-sample predictive accuracy. Applying this heuristic to our data, we see that almost all of the sub-periods where \GARCH appears to out-perform \RBGARCH have WAIC instabilities and that, ignoring those outliers, the \RBGARCH model consistently has superior in-sample performance, even after adjusting for the added complexity of the \RBGARCH model. 

A paired one-sided Wilcoxon test comparing the WAICs of the \GARCH and \RBGARCH models was performed, with the alternative hypothesis that the \RBGARCH model has a higher median WAIC across our 73 time periods (not omitting the outlier-corrupted periods). We see consistent evidence that \RBGARCH outperforms, with 37 out of 40 $p$-values less than 0.10 and 34 out of 40 less than 0.05. (Note that the non-parametric Wilcoxon test was used because the sampling distribution of WAIC is highly non-normal in small samples and extremely sensitive to the heteroscedasticity of the underlying data.) Overall, we see consistent evidence that, by jointly modeling the volatility of spot and futures returns, the \RBGARCH model is able to consistently capture price dynamics more accurately than the \GARCH model.

\subsection{In-Sample Risk Management}
\begin{figure}[ht]
  \centering
  \includegraphics[width=\textwidth,height=2.5in]{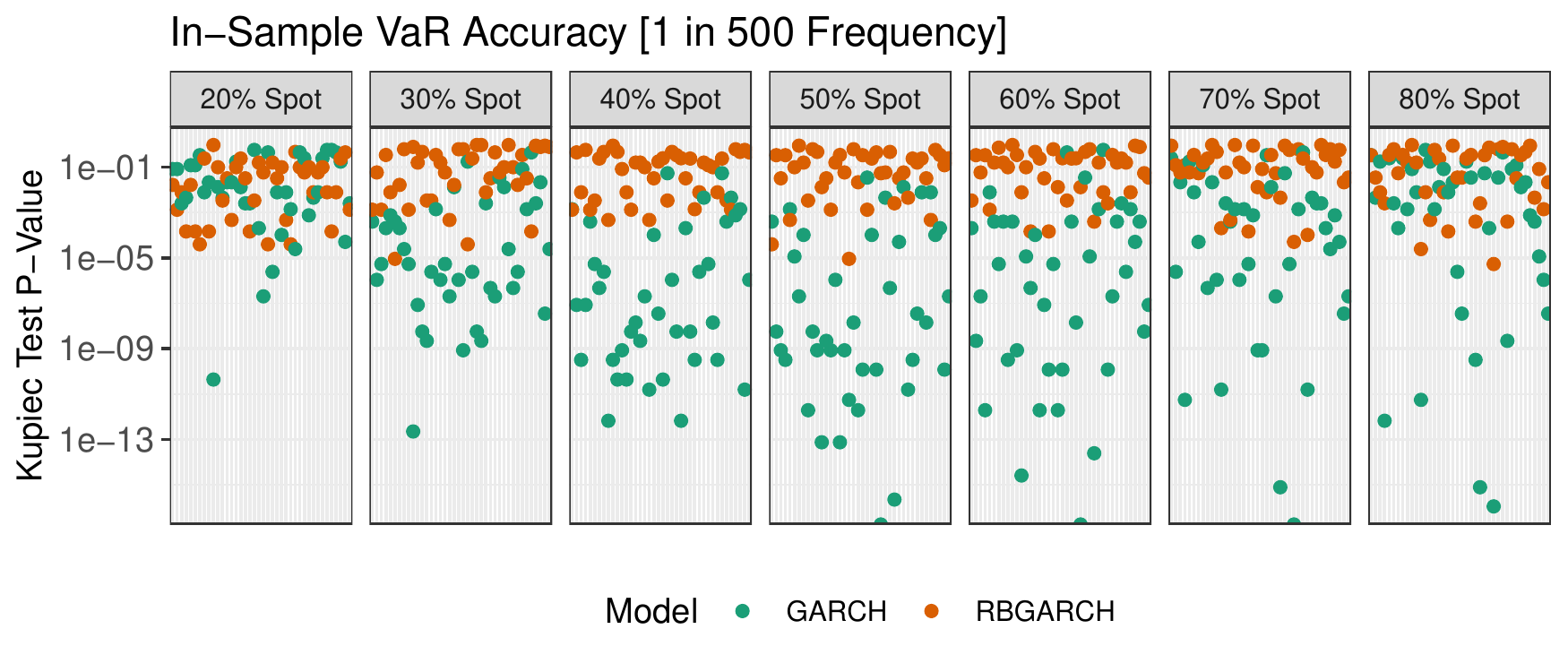}
  \caption{In-Sample VaR Accuracy of the \GARCH and \RBGARCH models, as measured by \citeauthor{Kupiec:1995}'s \citeyearpar{Kupiec:1995} unconditional exceedances test. Larger $p$-values indicate better performance. The \RBGARCH model performs well for all portfolios considered, while the \GARCH model shows clear signs of inaccurate VaR calculation for all portfolios except the 20\% Spot/80\% Futures portfolio. Displayed $p$-values are for in-sample VaR estimates for the entire sample period.}
  \label{fig:in_sample_var}
\end{figure}

We consider a multi-asset portfolio containing a mixture of spot LNG and one-month Henry Hub futures. In-sample VaR estimates are computed by taking quantiles of the posterior predictive distribution of portfolio returns using the posterior distribution of the mean return parameters $\mu, \alpha$ and the estimated conditional variance $\Sigma_t$. To assess the accuracy of our VaR estimates, we use the (unconditional) test of \citet{Kupiec:1995}, which compares the number of times that the actual losses are larger than the estimated quantile (VaR) with the theoretical rate using a binomial proportion test. The Kupiec test is a misspecification test, where smaller values indicate stronger evidence that the VaR estimates are not correctly calibrated. Results of this test are shown in Figure \ref{fig:in_sample_var}. As this figure shows, the \RBGARCH model consistently provides more accurate in-sample VaR estimates than the \GARCH model. We note that, because it models the volatilities jointly and does not assume independence, the \RBGARCH model does particularly well for balanced portfolios where the correlation among assets is particularly important. 

\subsection{Out-of-Sample Risk Management}
The ultimate test of a financial model is its out-of-sample performance, so we compare the out-of-sample performance of the \GARCH and \RBGARCH models, again focusing on VaR forecasting. Forward-looking (out-of-sample) volatility predictions were obtained by (forward) filtering observed returns for each posterior sample and then taking the quantile of the posterior predictive distribution for each day. In other words, out-of-sample observations were used to update the conditional forward-looking volatility estimates, but were not used for parameter re-estimation. While we only consider one-day ahead prediction here, it is straightforward to extend these results to longer forecast horizons using standard techniques, \emph{e.g.}, the data-driven aggregation method proposed by \citet{Hamidieh:2010}.

\begin{figure}[ht]
  \centering
  \includegraphics[width=\textwidth,height=2.5in]{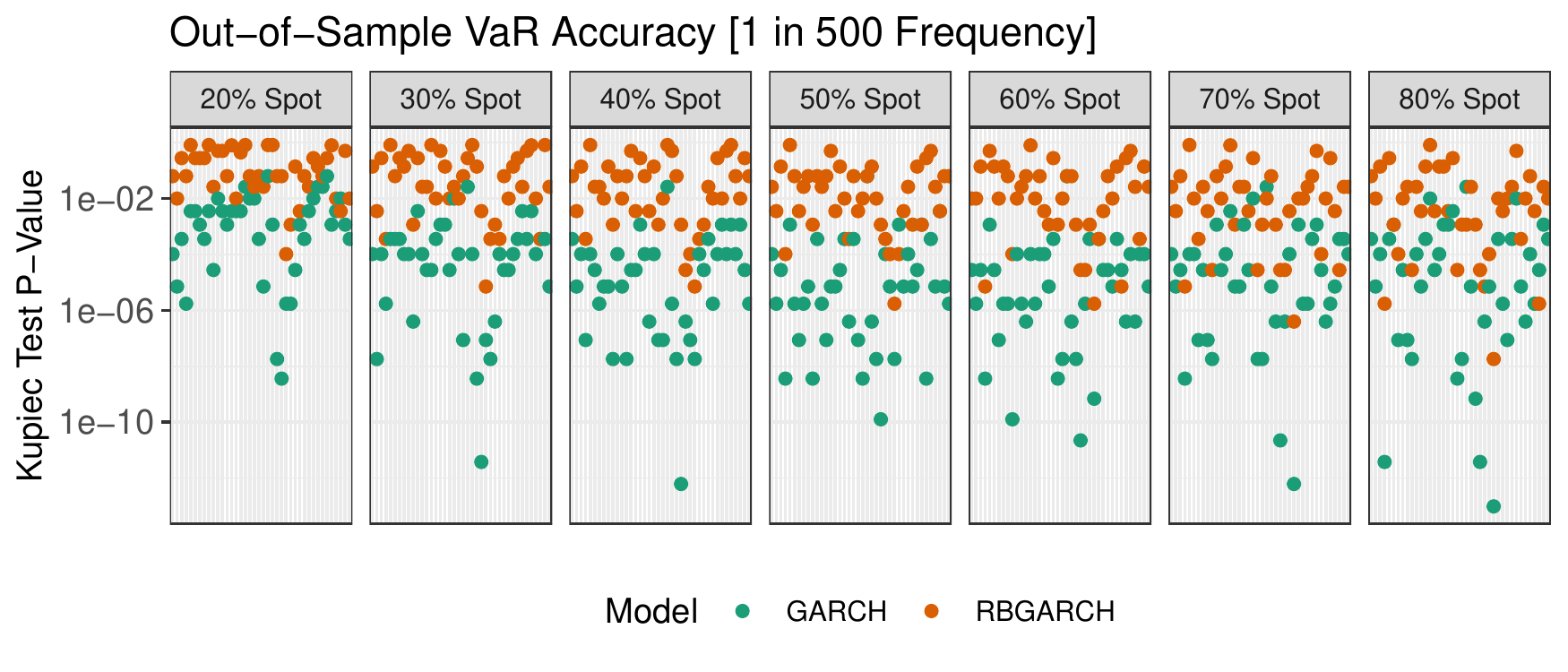}
  \caption{Rolling Out-Of-Sample VaR Accuracy of the \GARCH and \RBGARCH models, as measured by \citeauthor{Kupiec:1995}'s \citeyearpar{Kupiec:1995} unconditional exceedances test. Larger $p$-values indicate better performance. The \RBGARCH model outperforms the \GARCH model for the seven different portfolio weights considered here. Displayed $p$-values are for out-of-sample filtered VaR estimates for the entire sample period.}
  \label{fig:out_sample_var}
\end{figure}

Results of this test are shown in Figure \ref{fig:out_sample_var} for the same two-asset portfolios as considered in the previous section. The \RBGARCH model continues to consistently outperform the \GARCH model, particularly for the portfolios in which the spot asset is heavily weighted. In particular, while the performance of the \GARCH model is significantly worse out-of-sample, the \RBGARCH model performs almost as well in-sample as it does out-of-sample performance. A closer examination reveals that, while the \GARCH model significantly underestimates volatility, the \RBGARCH model produces slightly over-conservative out-of-sample VaR estimates, a feature which we consider acceptable, and possibly even desirable, in applications. 

\section{Conclusions} \label{sec:disc}
We present a model for joint volatility estimation at several LNG hubs and demonstrate that, despite the apparent mismatch, information from Henry Hub futures can be used to improve volatility estimation for non-Henry Hub spot prices. We provide suggested priors for use in Bayesian estimation and demonstrate the effectiveness of the Bayesian approach for accurate risk prediction. Our model produces improved volatility estimates, which exhibit more accurate estimates of unconditional volatility levels, increased responsiveness to external market shocks, and improved VaR estimation. Notably, our joint model performs as well out-of-sample as it does in-sample, making it particularly promising for risk management applications. This impressive out-of-sample performance can be traced back to our use of realized volatility information from the futures market: by taking advantage of an additional highly-accurate realized volatility measure, the Realized Beta GARCH is able to adapt to changing market conditions much more rapidly than standard approaches. Joint modeling of different aspects of LNG markets is an effective and flexible strategy for incorporating inconsistently available data and provides useful insights into the degree and dynamics of LNG spot price volatility and inter-asset correlation. The joint modeling framework may be easily extended to include additional realized volatility measures or to allow more complex market dynamics. 

\subsection*{Acknowledgements}

We thank the El Paso Corporation Finance Center at the Jones Graduate School of Business at Rice University and the Rice Center for Computational Finance and Economic Systems (CoFES) for assistance acquiring historical spot and futures price data. We also thank CoFES for providing computational resources. MW acknowledges support from NSF Graduate Research Fellowship Program under grant number 1842494. The authors have no conflicts of interest to declare.

\subsection*{Supplemental Materials}

Supplemental Materials to be published alongside the online version of this paper include additional descriptive analysis of our sample data, more details of our in- and out-of-sample performance, code necessary to replicate our data set, and a more detailed discussion of computational aspects of our approach, including the \texttt{Stan} code implementing Model \eqref{eqn:model}.

\printbibliography

\end{refsection}

\clearpage
\begin{refsection}
\appendix
\singlespacing

\setcounter{table}{0}
\renewcommand{\thetable}{A\arabic{table}}
\setcounter{figure}{0}
\renewcommand{\thefigure}{A\arabic{figure}}

\begin{center} {\huge \bf Supplemental Materials} \end{center}

\section{Additional Data Description} \label{app:descrip}
\begin{figure}
\begin{center}
\includegraphics[width=\textwidth]{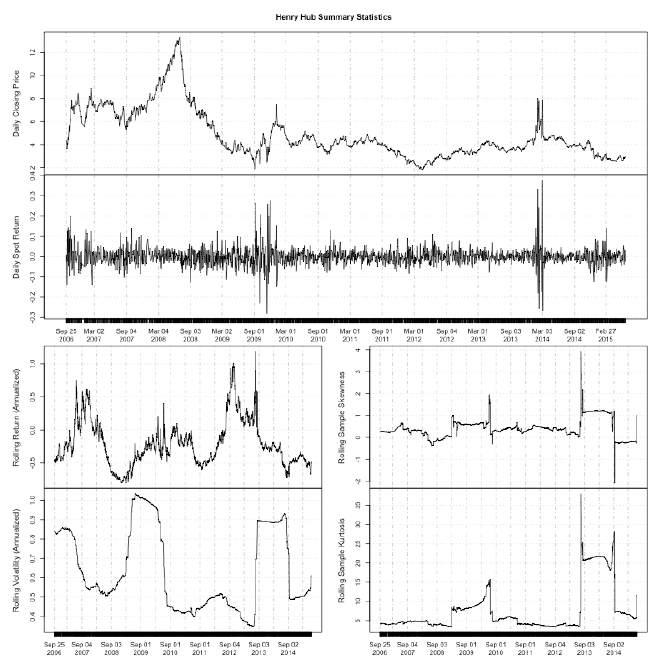}
\end{center}
\caption{Henry Hub spot prices and rolling estimates of the first four moments. Spot prices peaked in mid-2008, declined from late 2008 to early 2009 and have remained at a relatively constant level since. Spot returns exhibit moderate heteroscedasticity, skewness and kurtosis, with large changes in the sample moments being primarily driven by high-volatility periods in Fall 2009 and Spring 2014.}
\label{fig:hhub_spot}
\end{figure}

In this section, we provide some additional characterization of the data used for the empirical study in Section \ref{sec:application} of our paper. As discussed in Sections \ref{sec:mkts} and \ref{sec:data}, Henry Hub spot prices play an important role in both the spot and futures markets. Figure \ref{fig:hhub_spot} shows the spot price of Henry Hub futures for the sample period, as well as (annualized) rolling return and volatility estimates, sample skewness, and sample (non-excess) kurtosis. As \citet{Li:2019} notes our sample period is characterized by relatively low volatility, with changes in volatility, skewness, and kurtosis being driven by periods of elevated volatility in late 2009 and in early 2014. Given this low level of volatility, we do not include a jump component in our model, contrary to the findings of \citet{Mason:2014} who find the presence of a jump component significantly improves model fit in an earlier, more volatile, period.

\begin{figure}
\begin{center}
\includegraphics[width=\textwidth]{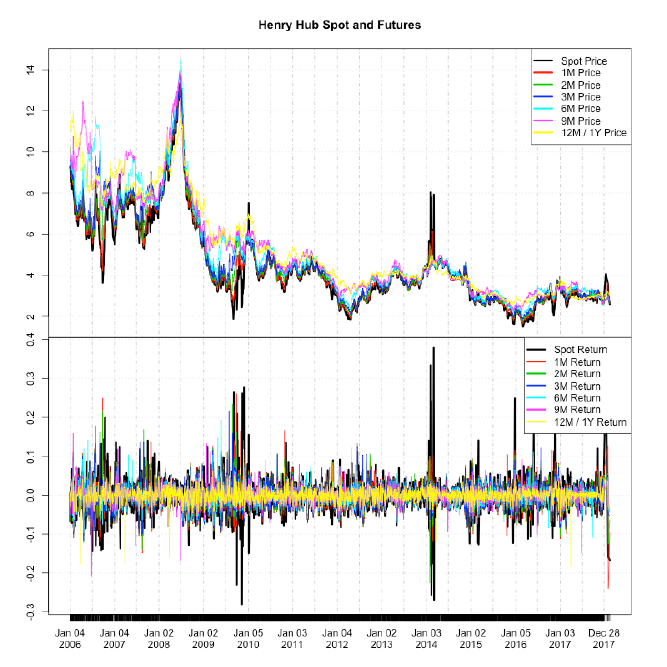}
\end{center}
\caption{Henry Hub futures move in tandem with spot prices, but generally exhibit fewer large jumps. In particular, the futures markets do not exhibit the large jumps in early 2014 that are a significant driver of the elevated kurtosis observed in spot prices during the same period.}
\label{fig:hhub_futures}
\end{figure}

\begin{figure}
\begin{center}
\includegraphics[width=\textwidth]{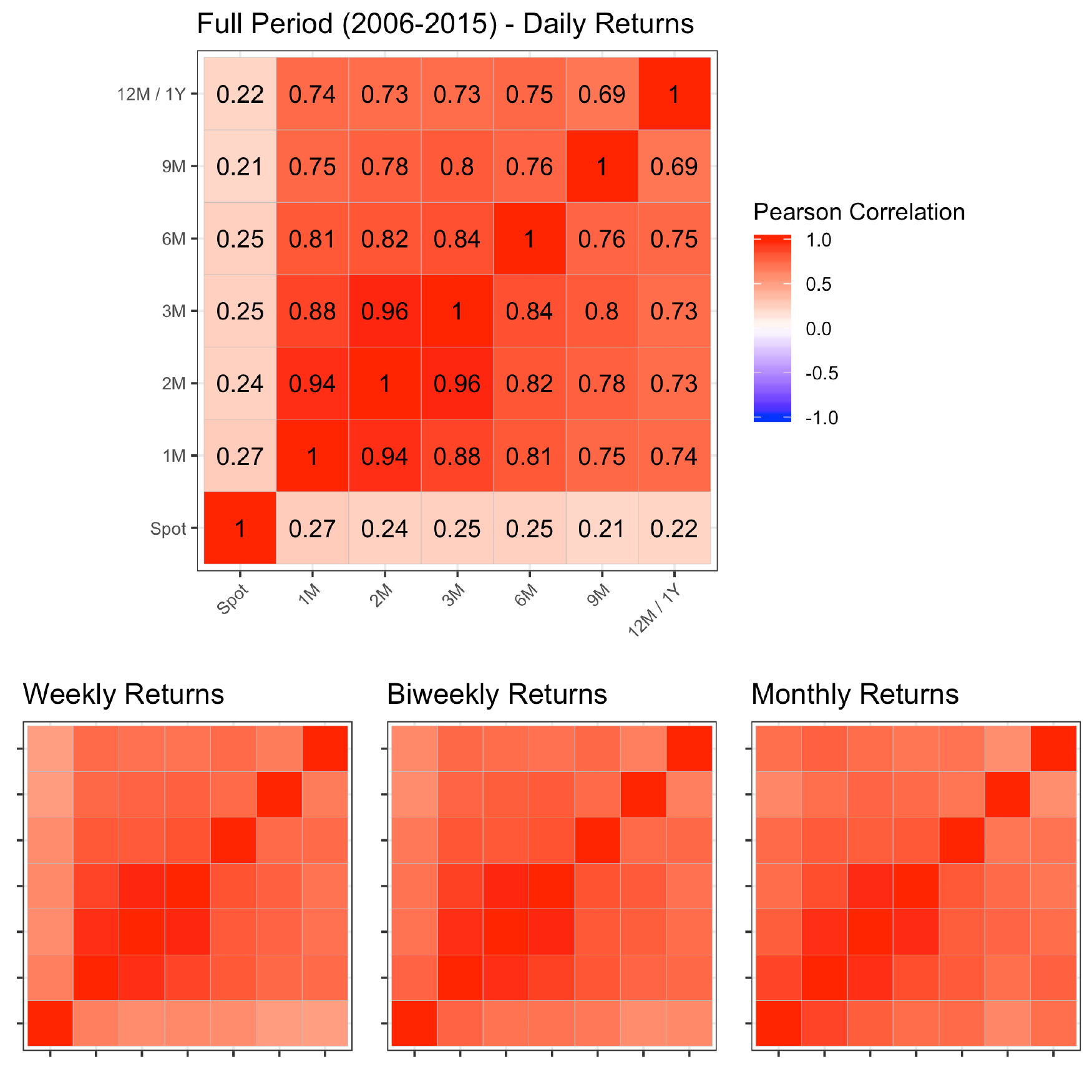}
\end{center}
\caption{Daily returns of Henry Hub futures are highly correlated, but only exhibit moderate correlation with Henry Hub spot returns. Correlations among spot and futures returns are higher at lower frequencies (weekly, monthly, \emph{etc.}).}
\label{fig:hhub_futures_cor}
\end{figure}

LNG futures prices are strongly correlated across different maturities, with the correlation remaining roughly constant across time as shown in Figure \ref{fig:hhub_futures}. Futures prices exhibit less pronounced jump behavior than the underlying spot price, particularly in the latter portion of our sample, leading to reduced correlation between future and sport return at a daily frequency. At lower frequencies, the effects of these jumps are reduced and the correlation between spot and futures returns is higher, as shown in Figure \ref{fig:hhub_futures_cor}. 

\begin{figure}
\begin{center}
\includegraphics[width=\textwidth]{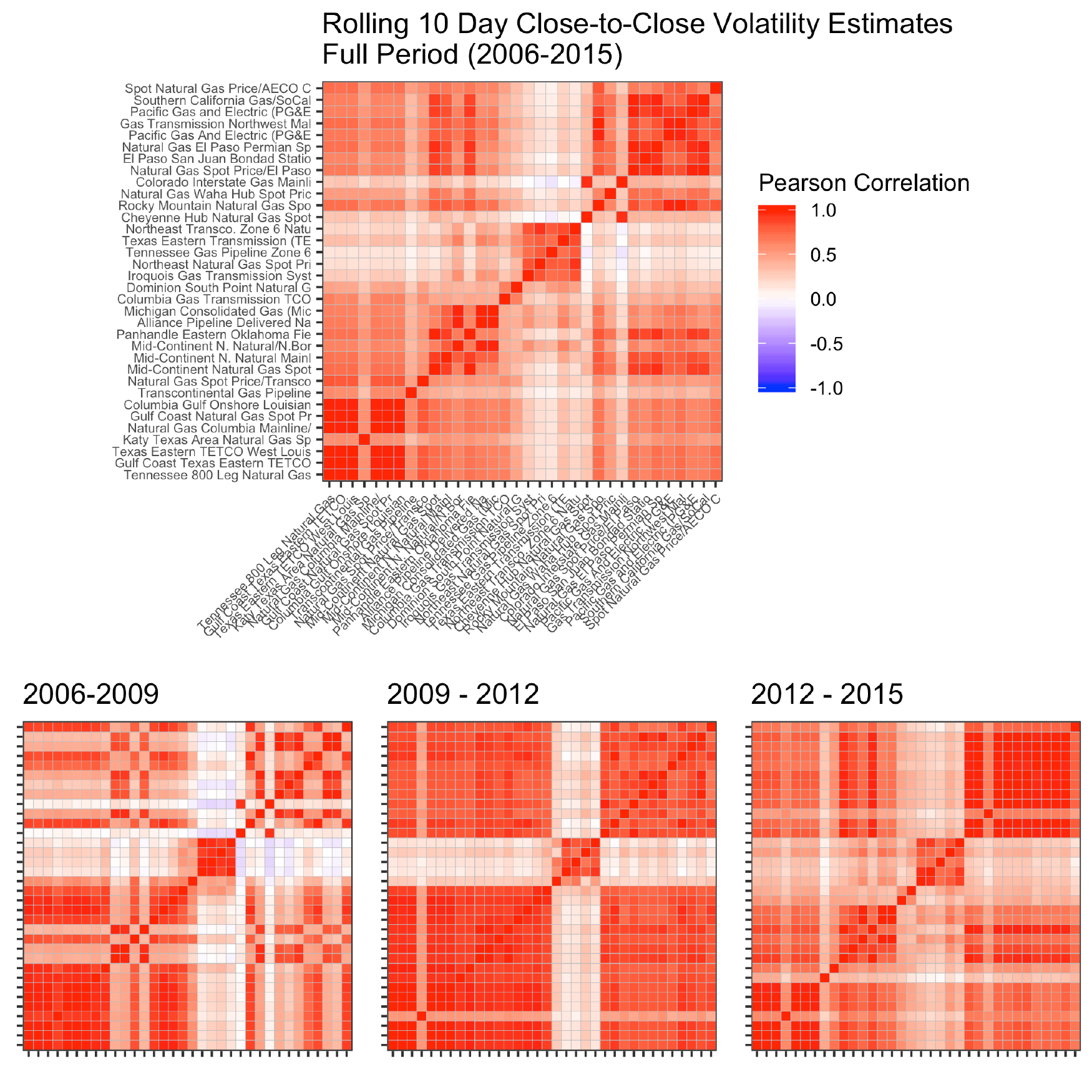}
\end{center}
\caption{Pearson correlation among the estimated volatilities of 40 spot prices during the sample period. As we see, there is generally a strong positive correlation among spot volatilities, though it is not as pronounced as the correlation among spot returns. There is some evidence that the structure of volatility is changing over time, as can be seen by the emergence of subgroups in the 2012-2015 period.}
\label{fig:spot_vol}
\end{figure}

Given that the effects of storage and production shocks are felt throughout LNG markets, we would expect  the \emph{volatilities} of spot prices are correlated as well. As can be seen in Figure \ref{fig:spot_vol}, this is indeed the case -- spot price volatilities are typically highly correlated, though not to quite the same degree as spot price returns. As before, there is clear evidence of a subgroup of spot prices that are more closely correlated among each other than to the rest of the market.

\begin{figure}
\begin{center}
\includegraphics[width=\textwidth]{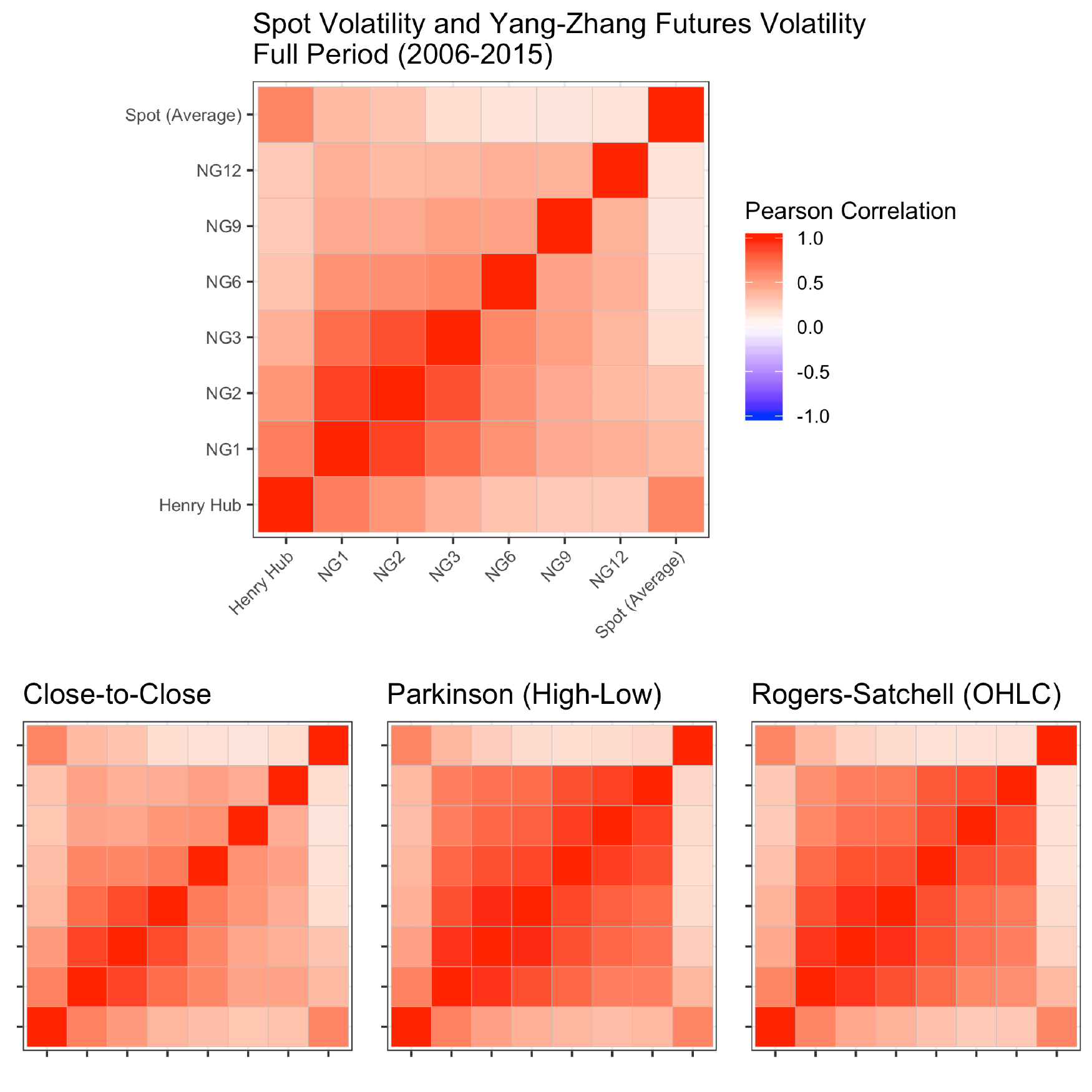}
\end{center}
\caption{Pearson correlation among the Henry Hub spot volatility, non-Henry Hub spot volatility, and futures realized volatility during the sample period. The reported correlations with non-Henry spots are the average correlation over the 40 non-Henry spots in our sample. The correlation between Henry Hub volatility and short-maturity (NG1) realized volatility is consistently the highest. The degree of correlation appears robust to the choice of realized volatility measure.}
\label{fig:yz_vol}
\end{figure}

While the spot and futures markets generally move together, they are still somewhat disjoint, with shocks to the futures market not necessarily being reflected in the spot market and \emph{vice versa}. Despite this disjoint behavior, there is still a strong correlation between realized volatility in the futures markets and close-to-close volatility in the spot markets. Figure \ref{fig:yz_vol} illustrates this correlation, using several different realized volatility measures and futures maturities. Henry Hub clearly displays a higher level of correlation with futures volatility than the other spot prices in our sample, consistent with the fact that LNG futures are based on Henry Hub prices. Regardless, futures volatility is indeed a useful predictor of volatility at non-Henry spots, as demonstrated in Section \ref{sec:application}.

The observed correlation does not appear to be sensitive to the choice of realized volatility measure, suggesting our results are robust to the exact form of external information being supplied. Shorter maturity futures, especially one-month futures (NG1), consistently show the highest correlation with spot prices, so we use the Yang-Zhang estimator applied to the generic fixed-maturity NG1 in our application. The high-degree of correlation suggests that single-factor volatility models, including our proposed model, will be able to capture most of the second-order dynamics. This is consistent with the findings of \citet{Kyj:2009} who, like us, combine a single-factor model with realized volatility measures and find significant improvements in equity portfolio construction.

\section{Data Set Replication Information} \label{app:series}
Table \ref{tab:series} lists the Bloomberg identifiers used to obtain the data for our analysis. For spot prices, only the last traded price of the day (Bloomberg key \texttt{PX\_LAST}) was used. These spot prices were selected from Bloomberg's full listing of LNG spot prices (available through the \texttt{BGAS} screen) as having a relatively complete trading history (more than 245 trading days per year) for the entire sample period (2006 - 2015). Days when only certain spots traded are omitted from our analysis to ensure an aligned data set. After screening, the \texttt{NGCDAECO} spot price (Alberta, Canada) was manually removed as it is the same spot as \texttt{NGCAAECO}, but reported in Canadian dollars instead of U.S.\, dollars.  For futures prices, Open, High, Low, and Close prices were obtained using Bloomberg keys \texttt{PX\_OPEN}, \texttt{PX\_HIGH}, \texttt{PX\_LOW}, and \texttt{PX\_CLOSE} respectively. 

As described in the main text, the reported Open and Close prices are occasionally outside of the daily trading range (the High-Low range). This is not a data error, but is instead an artifact of the Open and Close prices being set by an auction process which is not included in the High and Low calculations, rather than the standard market mechanisms. Despite this, the Yang-Zhang volatility estimator is derived assuming a continuous time model for the price series and makes no distinction between traded and non-traded prices. As such, it is not well-defined when the Open or Close prices fall outside of the intraday trading range. This occurs somewhat regularly in our data set, as detailed in Table \ref{app:tab:inconsistencies}, particularly at longer maturities. To address this, we restrict the Open and Close prices to the intraday trading range. This potentially results in a small loss of estimation efficiency, but empirically appears to have minimal effect, particularly for our application (Section \ref{sec:application}), where only the one month futures are used.

\begin{table}[hb]
\centering
\scalebox{0.8}{
\begin{tabular}{c|ccccc|c}
\toprule
& {\bf Open $>$ High} & {\bf Close $>$ High} & {\bf Open $<$ Low} & {\bf Close $<$ Low} & {\bf Low $>$ High} & {\bf Total}\\
\midrule
{\bf 1 Month (NG1)} & 6 & 0 & 5 & 0 & 0 & 11 \\
{\bf 2 Month (NG2)} & 3 & 1 & 2 & 1& 0 & 7\\
{\bf 3 Month (NG3)} & 9 & 4 & 7 & 0 & 0 & 20\\
{\bf 6 Month (NG6)} & 9 & 38 & 6 & 15 & 0 & 68\\
{\bf 9 Month (NG9)} & 11 & 121 & 9 & 61 & 0 & 202\\
{\bf 12 Month (NG12)} & 8 & 181 & 7 & 125 & 0 & 321\\
\midrule 
\textbf{Total} & 46 & 345 & 36 & 202 & 0 & 629\\
\bottomrule
\end{tabular}}
\caption{Apparent Inconsistencies in Futures Data. These typically occur because the auction-based Open and Close prices are not included in the intra-day trading range used to determine the High and Low prices. To calculate the Yang-Zhang \citeyearpar{Yang:2000} realized volatility, the Open and Close are truncated to the High-Low range.}
\label{app:tab:inconsistencies}
\end{table}

\begin{figure}
\begin{center}
\includegraphics[width=\textwidth]{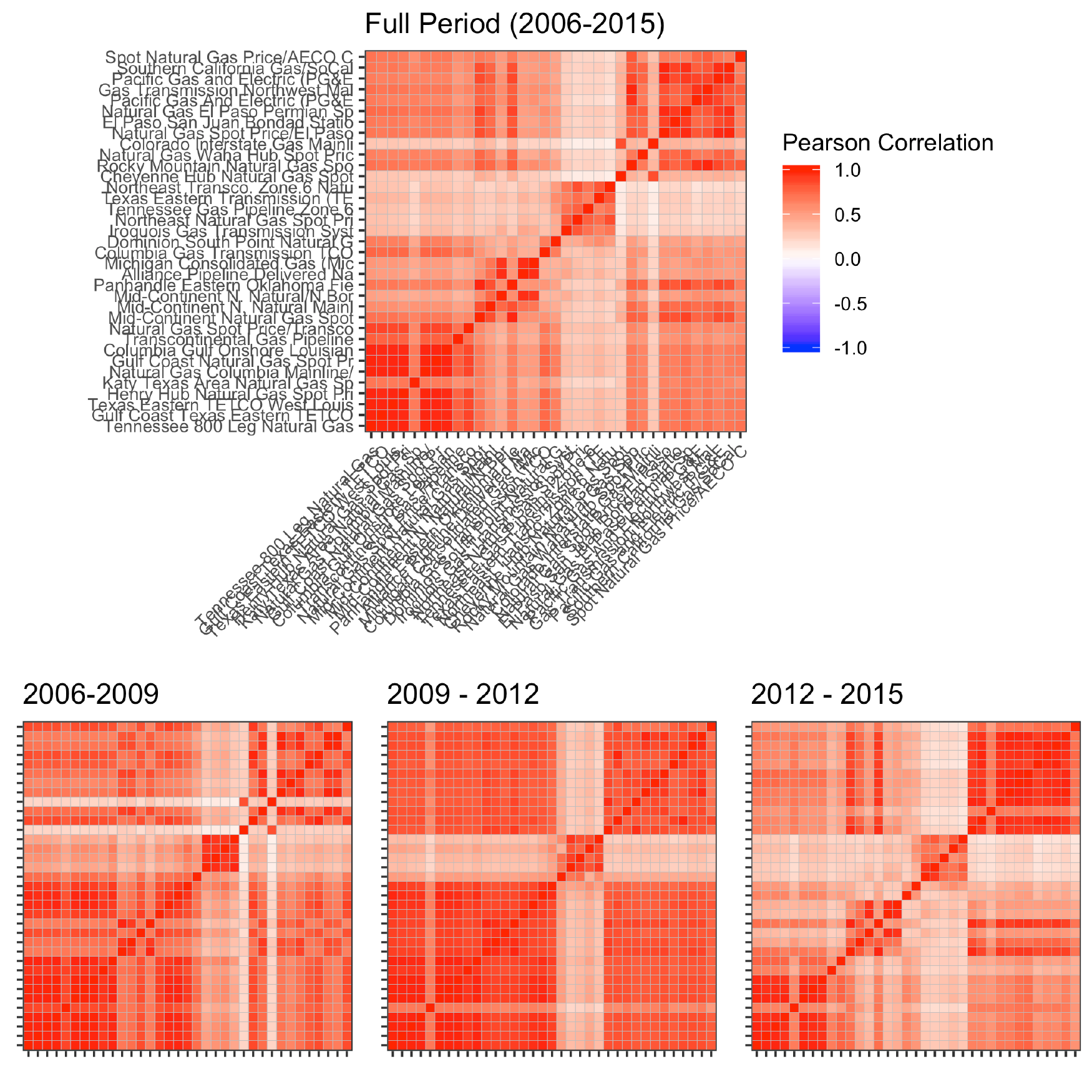}
\end{center}
\caption{Pearson correlation among daily spot price returns for 40 hubs in the sample period. Spot prices are typically highly correlated with each other and with Henry Hub (not shown here). The Cheyenne, Northeast Transco Zone 6, Texas Eastern Transmission, Tennessee Gas Pipeline, Northeast, and Iroquois Transmission System spot prices, however, are more highly correlated among themselves than to other spots during the sample period.}
\label{fig:corrgram}
\end{figure}
\clearpage

\begin{table}[phtb]
\centering
\footnotesize
\begin{tabular}{lll}
\toprule
{\bf Data Type} & {\bf Identifier} & {\bf Full Name}  \\ 
\midrule
\multirow{40}{*}{LNG Spot Prices} & NAGAALLI & Alliance Pipeline Delivered \\
& NAGAANRL & Mid-Continent / ANR Lavrne (Custer, OK) \\
& NAGAMICG & Michigan Consolidated Gas (Detroit, MI) \\
& NAGANGMC & Mid-Continent Natural Gas Spot Price \\
& NAGANGPL & Mid-Continent / Chicago Citygate (Chicago, IL) \\
& NAGANGTO & Texas-Oklahoma East (Montgomery County, TX) \\
& NAGANOND & N. Natural Mainline (Clay County, KS) \\
& NAGANORB & N. Border Natural Gas Spot Price (Ventura, IA) \\
& NGCAAECO & AECO C Hub (Alberta, Canada) \\
& NGCGNYNY & TETCO M3 (New York, NY) \\
& NGGCANRS & Gulf Coast / ANR Southeast \\
& NGGCCGLE & Columbia Transmission TCO Pool (Leach, KY) \\
& NGGCCOLG & Columbia Gulf Onshore Louisiana Pool \\
& NGGCT800 & Tennessee 800 Leg (SE Louisiana) \\
& NGGCTR30 & Transco Station 30 (Wharton County, TX) \\
& NGGCTRNZ & Transco Station 65 (Beauregard Parish, LA) \\
& NGGCTXEW & TETCO West Louisiana \\
& NGGCTXEZ & TETCO East Louisiana \\
& NGNECNGO & Dominion South Point (Lebanon, OH) \\
& NGNEIROQ & Iroquois Gas Transmission System (Waddington, NY) \\
& NGNEIZN2 & Iroquois Zone 2 (Wright, NY) \\
& NGNETNZ6 & Tennessee Gas Pipeline Zone 6 \\
& NGNETRNZ & Northeast Transco Zone 6 (Linden, NJ) \\
& NGRMCHEY & Cheyenne Hub (Cheyenne, WY) \\
& NGRMDENV & Colorado Interstate Gas Mainline \\
& NGRMELPS & Non-Bondad San Juan Basin (El Paso, TX / Blanco, NM) \\
& NGRMEPBD & Bondad San Juan Basin (El Paso, TX) \\
& NGRMKERN & Rocky Mountains / Kern River (Opal, WY) \\
& NGRMNWES & Northwest Pipeline (Stanfield, OR) \\
& NGTXEPP2 & Permian Basin (West Texas) \\
& NGTXOASI & Waha Hub (Waha, TX) \\
& NGTXPERY & Columbia Mainline (Perryville, LA) \\
& NGUSHHUB & Henry Hub (Erath, LA) \\
& NGWCEPEB & El Paso South Mainline (El Paso, TX) \\
& NGWCPGNE & Pacific Gas and Electric Citygate (N. California) \\
& NGWCPGSP & Northwest Transmission (Malin, OR) \\
& NGWCPGTP & Pacific Gas and Electric Topock (Topock, AZ) \\
& NGWCSCAL & Southern California Border \\
& NTGSTXKA & Katy Texas Area \\
& SNNWPIPA & Eastern Oklahoma Panhandle Field Zone (Haven, KS) \\
\midrule
\multirow{6}{*}{LNG Futures Prices} & NG1 & Generic 1st NG Future \\
 & NG2 & Generic 2nd NG Future \\
 & NG3 & Generic 3rd NG Future \\
 & NG6 & Generic 6th NG Future \\
 & NG9 & Generic 9th NG Future \\
 & NG12 & Generic 12th NG Future \\
\bottomrule
\end{tabular}
\caption{Bloomberg identifiers of the price series used in our analysis. LNG Futures were used to compute realized volatility measures and served as the market proxy while LNG Spot prices were used for the sparse-data asset.}
\label{tab:series}
\end{table}

\clearpage
\section{Additional Results} \label{app:additional_results}

In this section, we provide additional description of the in- and out-of-sample performance of our \RBGARCH model, extending the results presented in Section \ref{sec:application}. 

\subsection{In-Sample Model Fit}

We begin by considering the in-sample fit, as measured by the Widely Applicable Information Criterion (WAIC) proposed by \citet{Watanabe:2010}. As discussed in more detail by \citet{Gelman:2014}, WAIC is a fully Bayesian analogue of AIC \citep{Akaike:1974}, which attempts to estimate out-of-sample expected log-posterior density. Unlike the Deviance Information Criterion (DIC) of \citet{Spiegelhalter:2002,Spiegelhalter:2014}, WAIC is calculated using the entire posterior distribution, rather than just the posterior mean, taking fuller account of posterior uncertainty. Note that the Bayesian WAIC should not be confused with the so-called \emph{Bayesian} Information Criterion (BIC) of \citet{Schwarz:1978} or its actually Bayesian analogue WBIC \citep{Watanabe:2013}, both of which attempt to estimate the marginal likelihood of a proposed model \citep{Neath:2012}. Because we do not assume that either the \GARCH or \RBGARCH models are true, we focus more on predictive accuracy than model recovery.

\begin{figure}[ht]
  \centering
  \includegraphics[width=\textwidth]{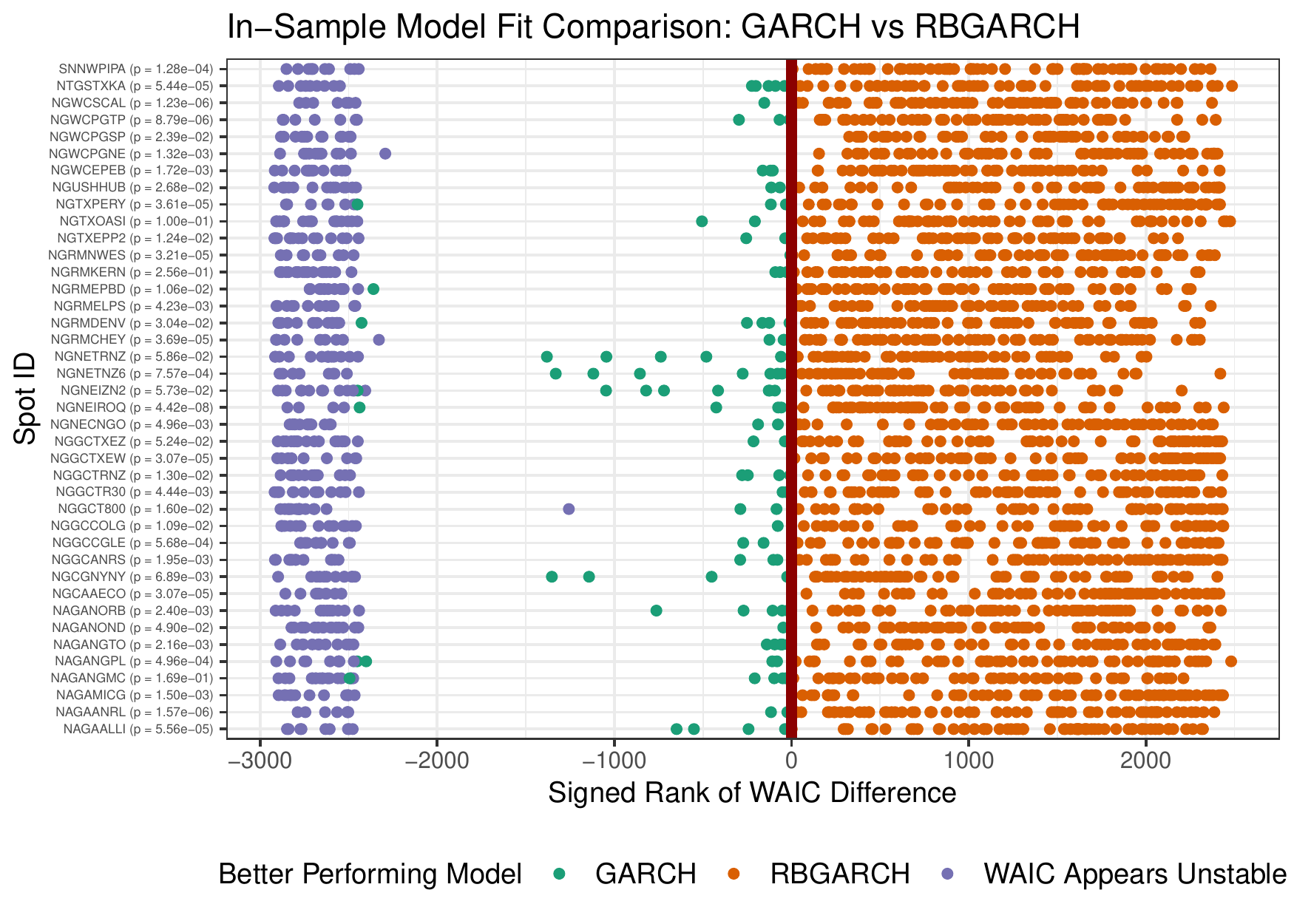}
  \caption{Signed-rank-transformed differences of 40 spots over 73 periods. Positive values indicate that the \RBGARCH model outperformed the \GARCH model for that sample period. $p$-values from a paired one-sample Wilcoxon test applied to WAIC differences for each sample are displayed on the left axis. The \RBGARCH model outperformed the \GARCH model across all spots, but was somewhat more sensitive to outliers in the data.}
  \label{fig:waic_app}
\end{figure}

\begin{figure}[htb]
  \centering
  \includegraphics[height=2in]{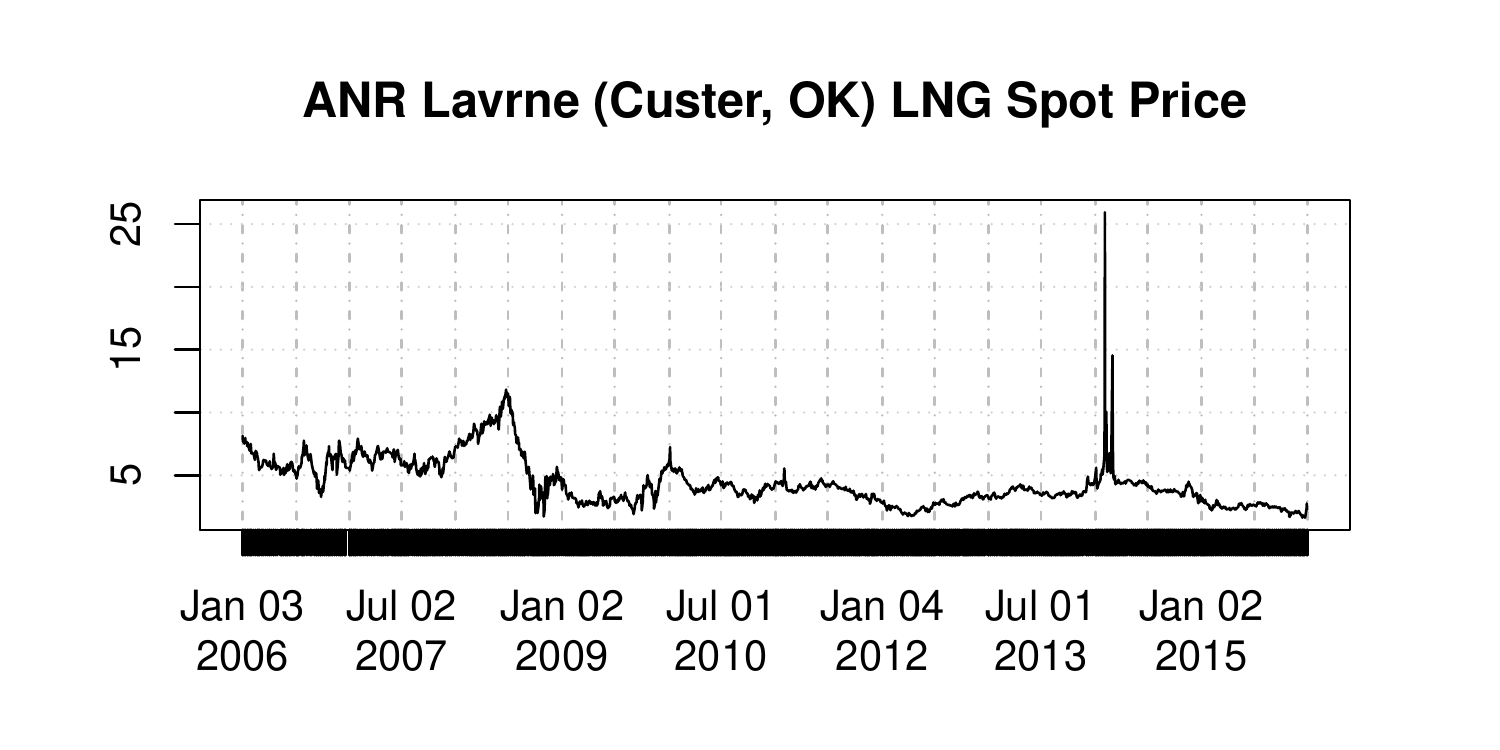}
  \caption{Price history for Custer, OK LNG trading hub. The large price jump on February 5\textsuperscript{th}, 2014 causes WAIC instabilities for several sample sub-periods, as highlighted in Figure \ref{fig:waic_app}. Because we use a 50-day rolling window with a 250-day history, a single outlier of this form can impact up to 5 sub-periods.}
  \label{app:fig:outlier}
\end{figure}

Figure \ref{fig:waic_app} shows in-sample WAIC comparisons of the \GARCH and \RBGARCH models for each of our 40 spot prices and 73 sub-sample periods. As discussed above, for a small fraction of sub-samples 5-15\%, depending on the spot, the WAIC estimates are corrupted by the presence of extreme outliers (see, \emph{e.g.}, Figure \ref{app:fig:outlier}). We identify these corrupted sub-periods using the 0.4 posterior variance heuristic of \citet{Vehtari:2017} and highlight them in purple. This heuristic identifies several periods in which the WAIC estimates of the \RBGARCH model are unstable, but those of the \GARCH model are not. We conjecture that this is due to the additional structure of the \RBGARCH model: because the \RBGARCH model expects volatilities to move together, an outlier in the spot price which does not have a corresponding jump in the futures market is even less likely under the \RBGARCH model than a simultaneous jump in both markets. Regardless, for our specific goal of VaR forecasting, the \RBGARCH model consistently outperforms the \GARCH model with no outlier filtering necessary, as shown in the following two sections.

\begin{figure}[htb]
  \centering
  \includegraphics[height=2.6in]{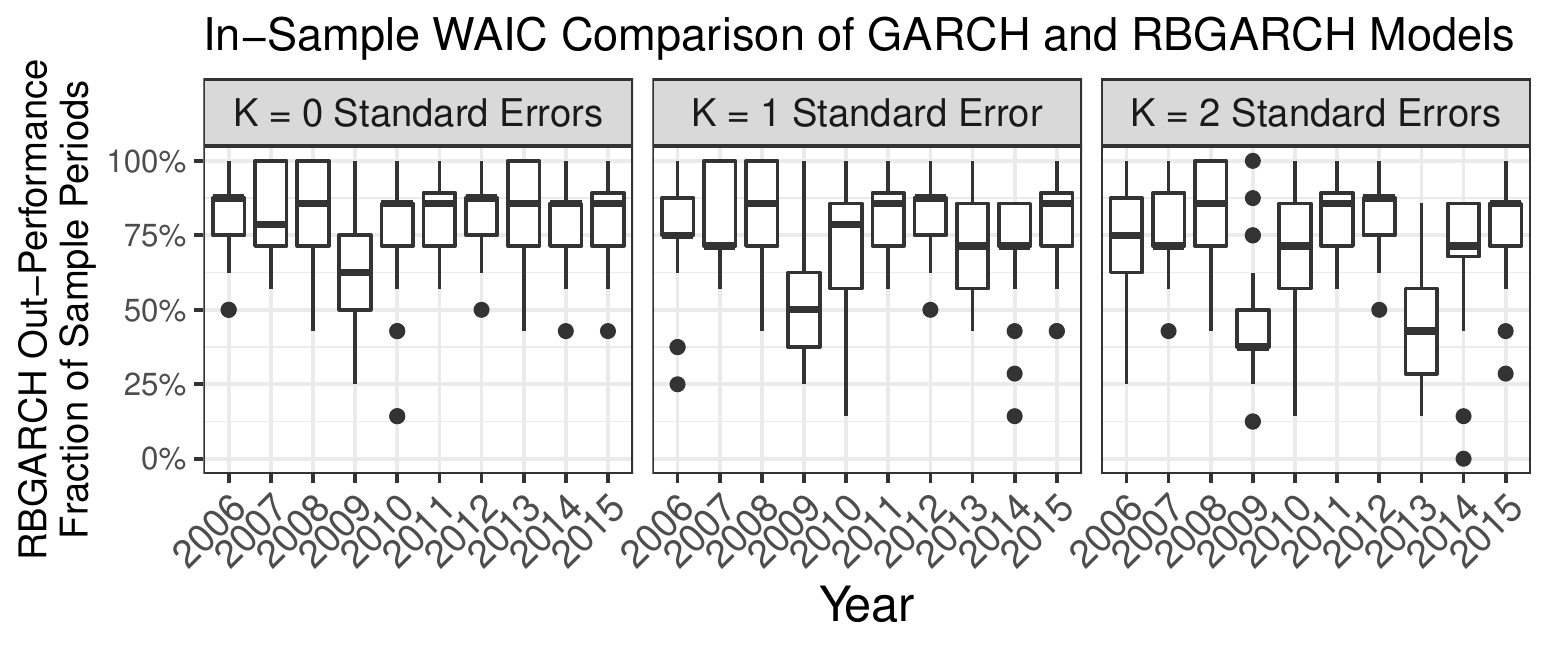}
  \caption{In-Sample WAIC Comparison of \GARCH and \RBGARCH models, aggregated over all 40 spot prices in our sample. The reported probabilities are the fraction of 250 day sample periods in which the the WAIC of the \RBGARCH exceeded that of the \GARCH model by at least $K$ times the estimated standard error of the difference for $K = 0, 1, 2$. For all years in the sample period, the \RBGARCH model has a consistently higher WAIC than the \GARCH model, typically by several standard errors.}
  \label{app:fig:waic}
\end{figure}

In Figure \ref{app:fig:waic}, we show the fraction of subsample periods for which the \RBGARCH model outperforms the \GARCH model by $K$ standard errors, broken down by year. (Because WAIC calculations include a standard error, it is possible to obtain a standard error on the difference in WAIC values. For details, see Section 5 of \citet{Vehtari:2017}.) The \RBGARCH model consistently outperforms the \GARCH model, typically by one or more standard errors, for all years in the sample period.

\subsection{In-Sample Risk Management}

\begin{figure}
  \centering
  \includegraphics[width=\textwidth]{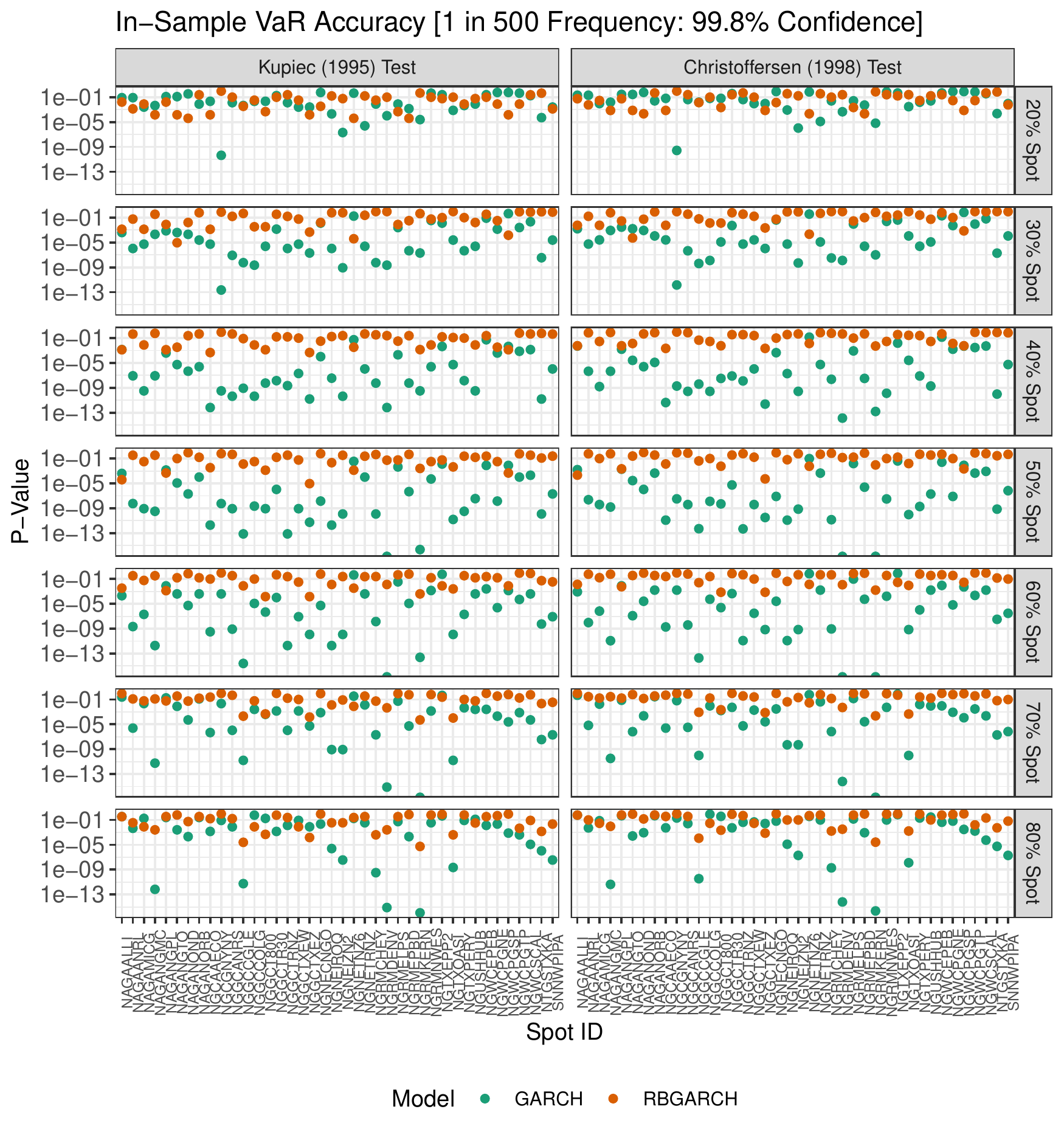}
  \caption{In-sample VaR accuracy measures of the \GARCH and \RBGARCH models at the 99.8\% (1 in 500 days) confidence level. The \citet{Kupiec:1995} and \citet{Christoffersen:1998} tests both indicate that the \RBGARCH model consistently outperforms the \GARCH model for all 40 spots in our sample, with the advantage being particularly pronounced for approximately equally weighted portfolios. In order to accurately evaluate VaR predictions at this extreme quantile, the reported $p$-values were obtained by taking the concatenated in-sample predictions over the entire 10-year sample period.}
  \label{app:fig:in_500}
\end{figure}

\begin{figure}
  \centering
  \includegraphics[width=\textwidth]{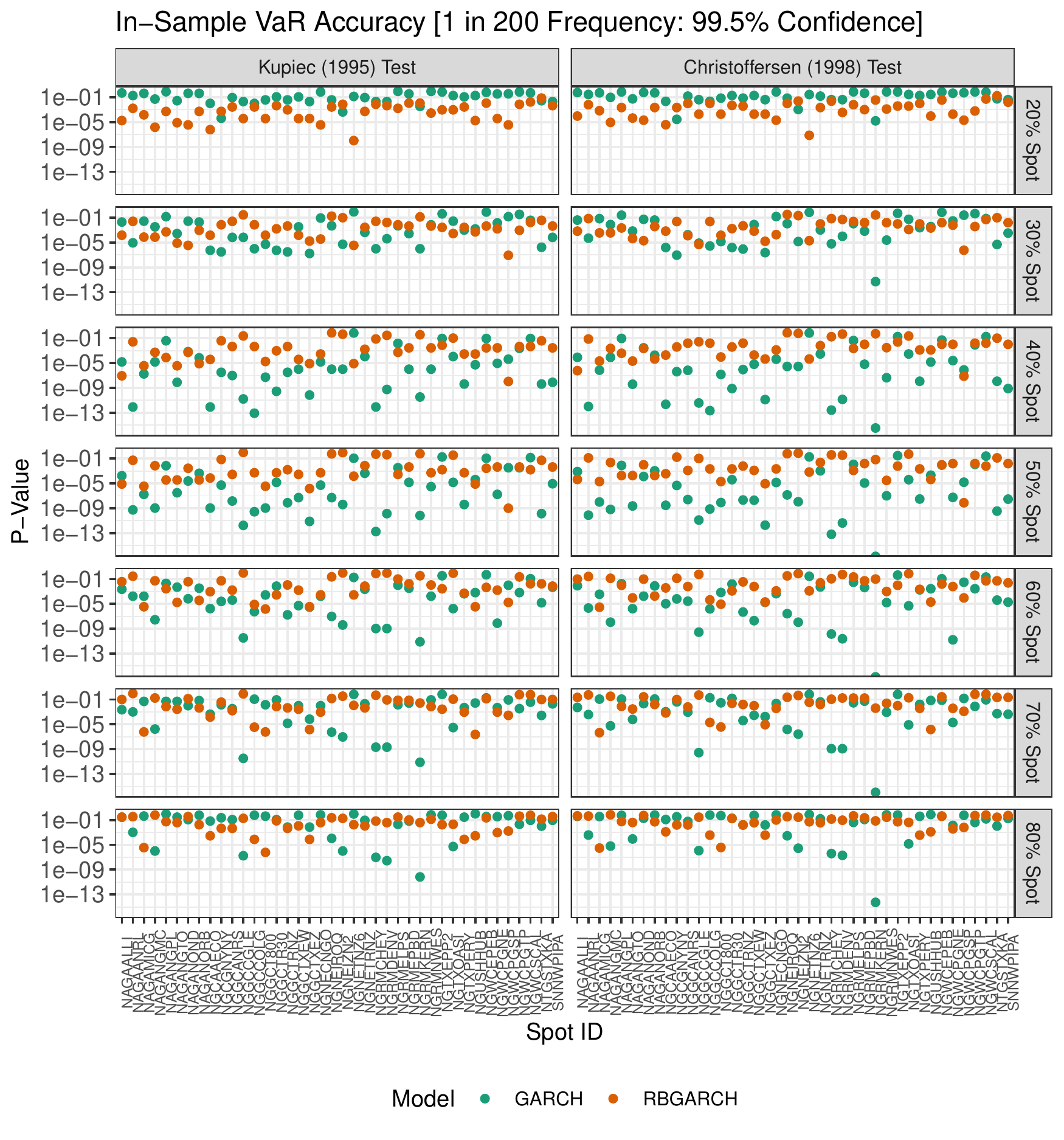}
  \caption{In-sample VaR accuracy measures of the \GARCH and \RBGARCH models at the 99.5\% (1 in 200 days) confidence level. Compared to Figure \ref{app:fig:in_500}, the in-sample advantage of the \RBGARCH model is still present, albeit less pronounced, as both models are able to capture this portion of the distribution well.}
  \label{app:fig:in_200}
\end{figure}

\begin{figure}[t]
  \centering
  \includegraphics[width=\textwidth]{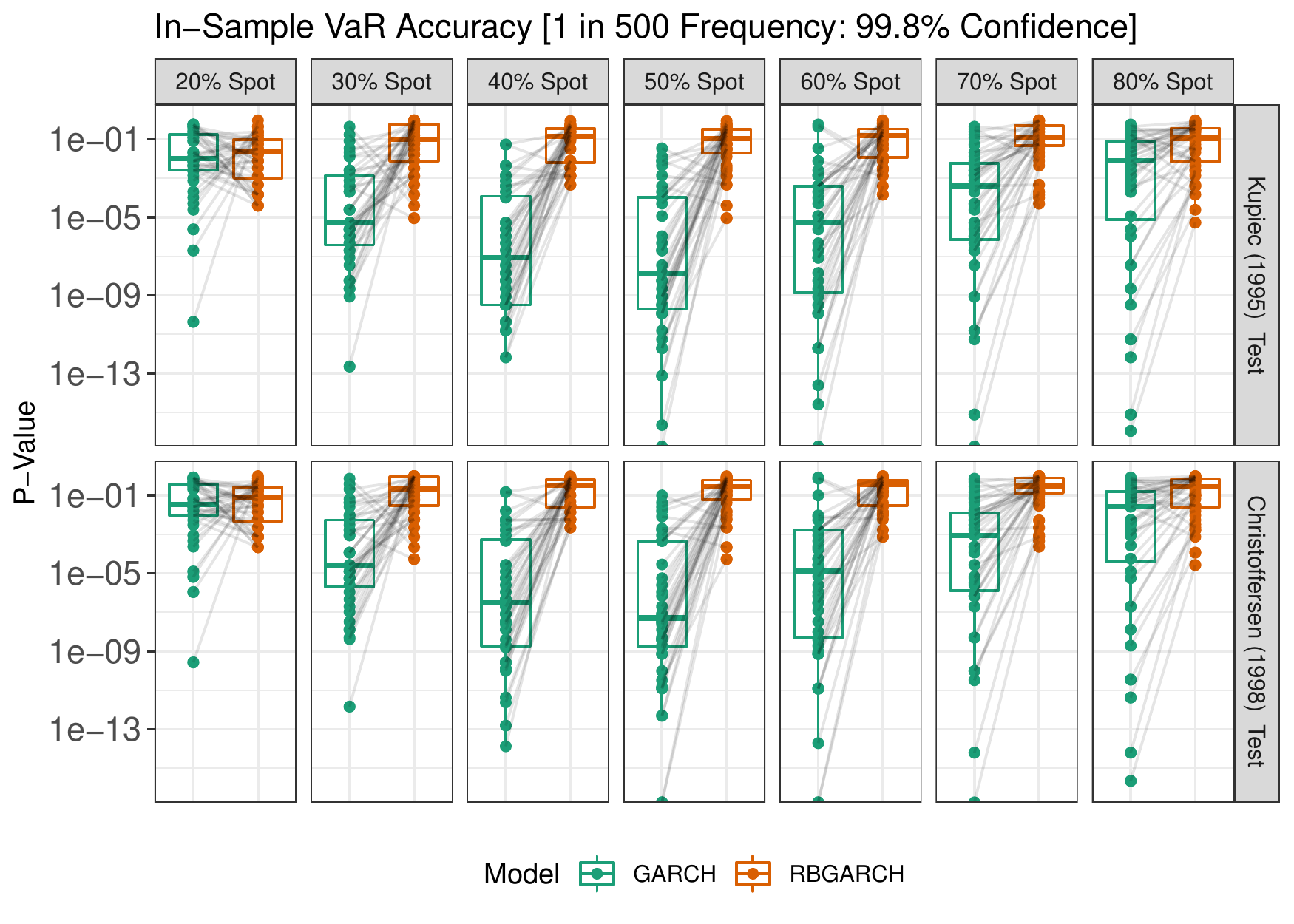}
  \caption{In-sample VaR accuracy measures of the \GARCH and \RBGARCH models at the 99.8\% (1 in 500 days) confidence level. Consistent with Figure \ref{app:fig:in_500}, the \RBGARCH model consistently outperforms the \GARCH model, as can be seen from the upward sloping lines connecting corresponding $p$-values from the two models.}
  \label{app:fig:in_var_paired}
\end{figure}

In addition to the unconditional test of \citet{Kupiec:1995}, we can also assess VaR accuracy using the conditional test of \citet{Christoffersen:1998}. While the Kupiec test is a marginal goodness of fit test, evaluating whether the samples appear to be draws from a Bernoulli distribution with nominal coverage, the Christoffersen test considers dependence between samples and tests whether the probability of an exceedance on day $T$ is independent of the probability of an exceedance on day $T + 1$. Figures \ref{app:fig:in_500} and \ref{app:fig:in_200} display $p$-values for each of the 40 spots, at 99.8\% and 99.5\% levels respectively. Figure \ref{app:fig:in_var_paired} gives a simplified version of this information, from which we can clearly see the improved performance of the \RBGARCH model, as indicated by the upward sloping lines comparing the $p$-values for each model. 

\begin{figure}[b]
  \centering
  \includegraphics[width=\textwidth]{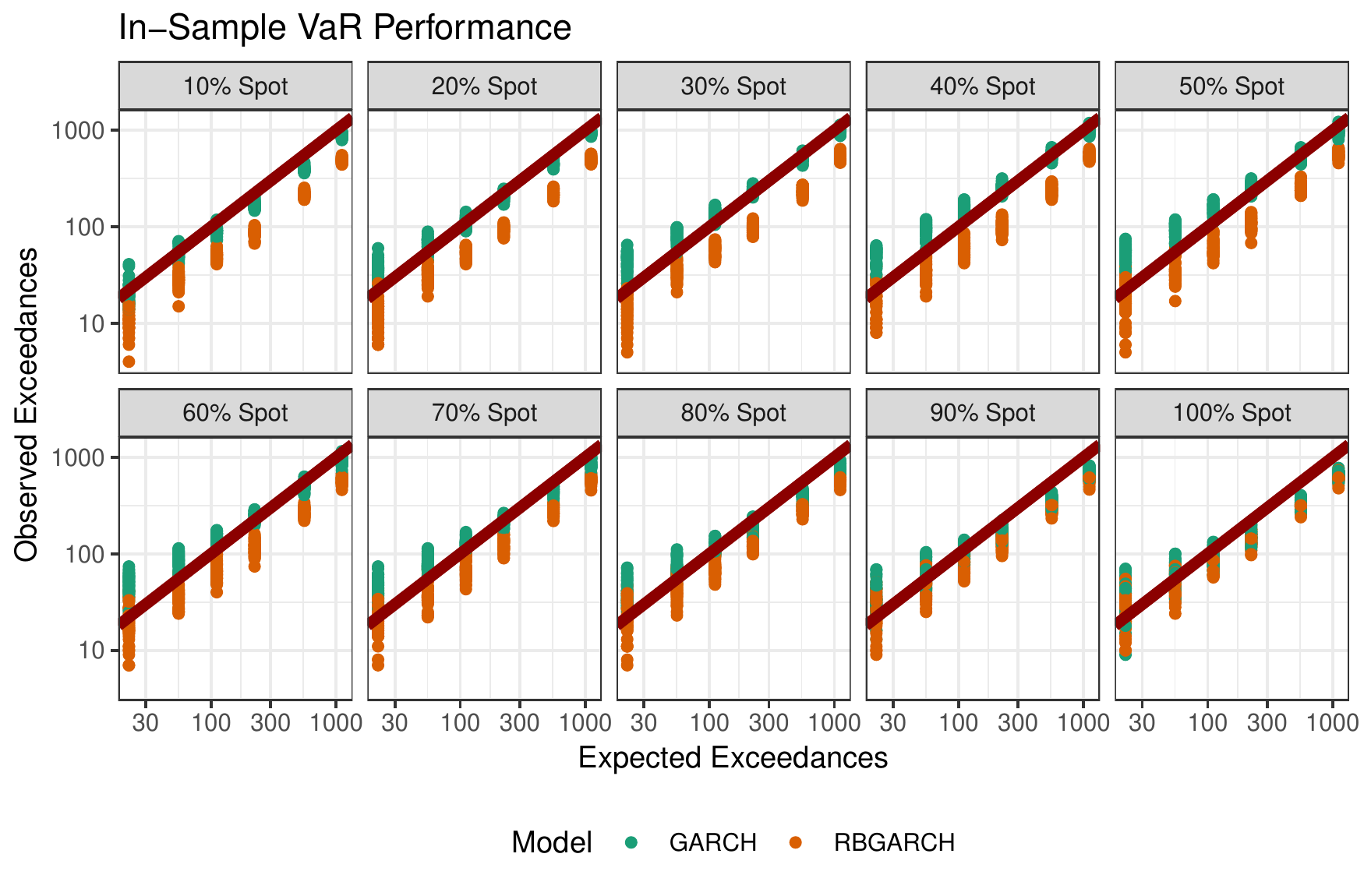}
  \caption{Assessment of the In-Sample VaR estimates of the \GARCH and \RBGARCH models for a range of sample portfolios. The \GARCH model typically underestimates the true VaR (lies above the red line) while the \RBGARCH model is more accurate, but with a conservative bias (lies below the red line).}
  \label{app:fig:in_sample_var}
\end{figure}

From these, we see that the \RBGARCH model consistently outperforms the \GARCH model, particularly for portfolios with approximately 50\% allocation in the spot and futures, taking advantage of its multivariate structure. While it is clear from these plots that the \RBGARCH model performs well, it is interesting to examine the limitations of the \RBGARCH model. To this end, Figure \ref{app:fig:in_sample_var} shows the expected and observed number of exceedances, corresponding to different confidence levels in the VaR calculation, for the \GARCH and \RBGARCH models. As can be seen from these figures, neither the \GARCH nor the \RBGARCH are perfectly calibrated (signified by the red line), though the two models fail differently. The \GARCH model typically has more observed exceedances than expected, indicating a systematic underestimation of VaR, while the \RBGARCH has generally fewer observed than expected, indicating overly conservative estimates. This is consistent with what we would expect for a well-fitting Bayesian forecast, as discussed in \citet{Ardia:2017}. (Note that the VaR estimates provided by both \GARCH and \RBGARCH approaches are more conservative than would be obtained from their non-Bayesian counterparts, but the \RBGARCH model is especially conservative as it accounts for parameter uncertainty in many more parameters than the \GARCH model.)

\subsection{Out-of-Sample Risk Management}

\begin{figure}
  \centering
  \includegraphics[width=\textwidth]{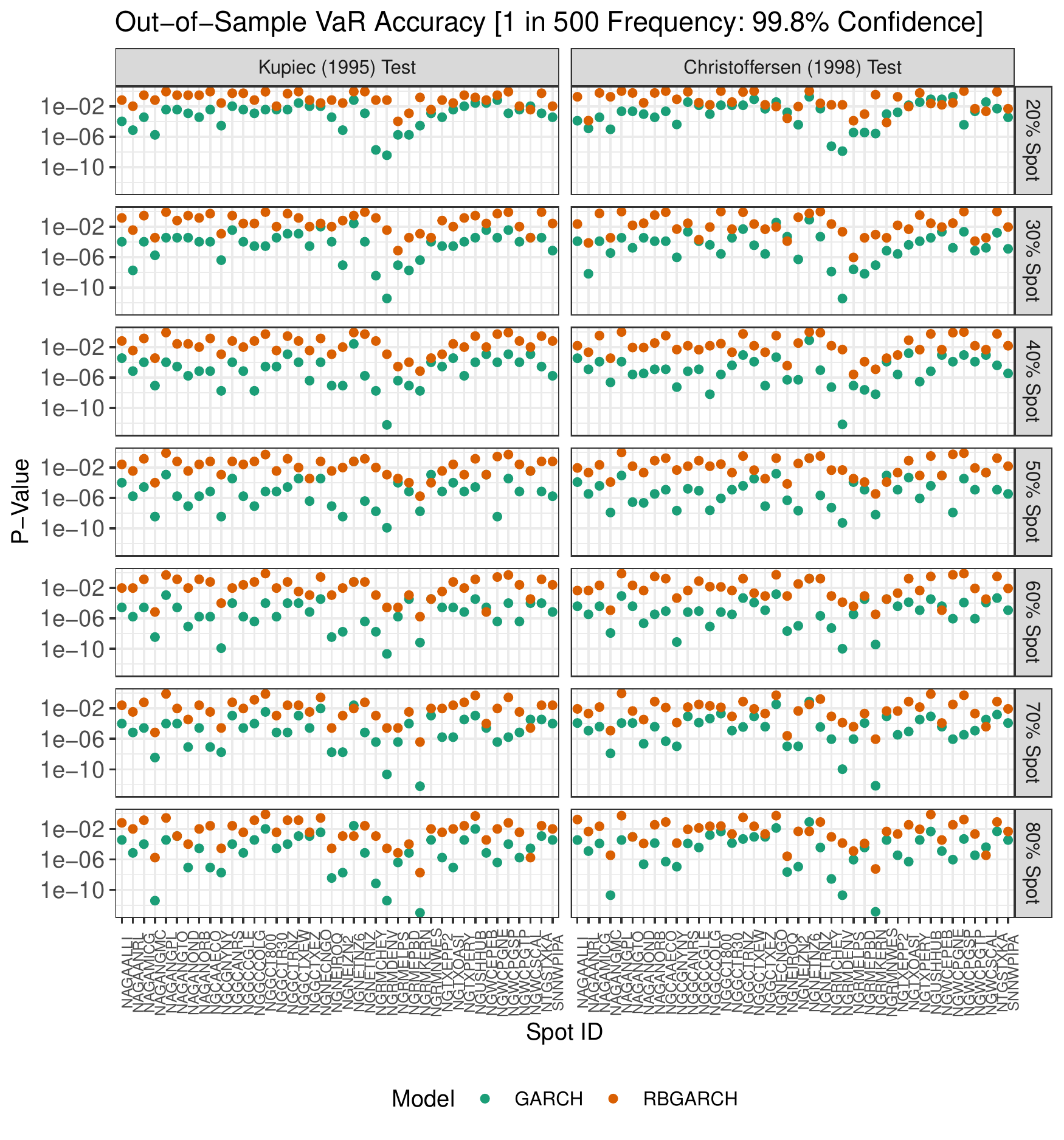}
  \caption{Out-of-sample VaR accuracy measures of the \GARCH and \RBGARCH models at the 99.8\% (1 in 500 days) confidence level. The \citet{Kupiec:1995} and \citet{Christoffersen:1998} tests both indicate that the \RBGARCH model consistently outperforms the \GARCH model for all 40 spots in our sample, with the advantage being particularly pronounced for approximately equally weighted portfolios. Somewhat surprisingly, the \RBGARCH model appears to perform almost as well out-of-sample as it does in-sample, while the \GARCH model is noticeably less accurate.}
  \label{app:fig:out_500}
\end{figure}

\begin{figure}
  \centering
  \includegraphics[width=\textwidth]{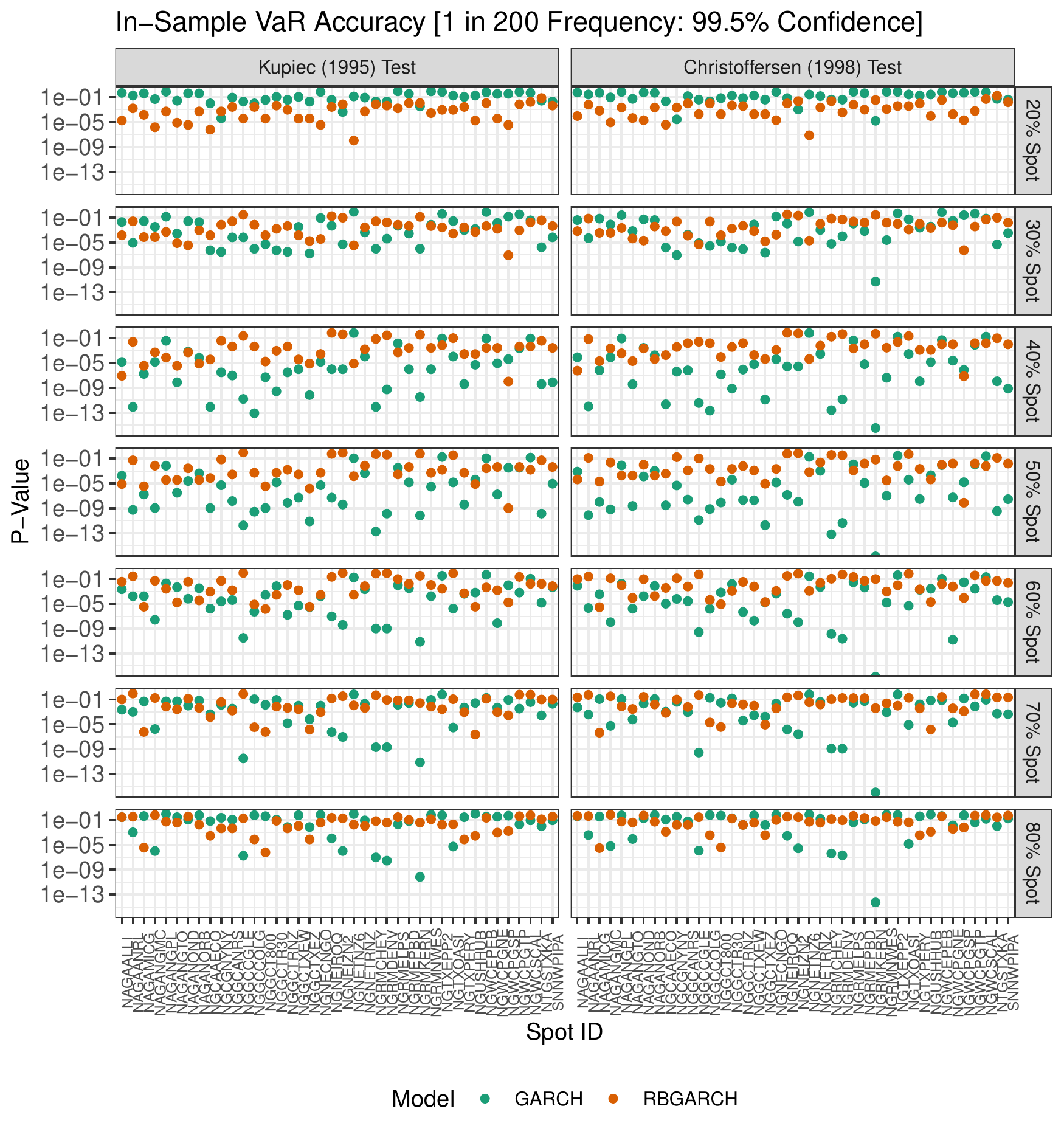}
  \caption{Out-of-sample VaR accuracy measures of the \GARCH and \RBGARCH models at the 99.5\% (1 in 200 days) confidence level. The \citet{Kupiec:1995} and \citet{Christoffersen:1998} tests both indicate that the \RBGARCH model consistently outperforms the \GARCH model for all 40 spots in our sample, with the advantage being particularly pronounced for approximately equally weighted portfolios. Unlike the in-sample case (Figure \ref{app:fig:in_200}), the \RBGARCH model here clearly outperforms the \GARCH model.}
  \label{app:fig:out_200}
\end{figure}

\begin{figure}[t]
  \centering
  \includegraphics[width=\textwidth]{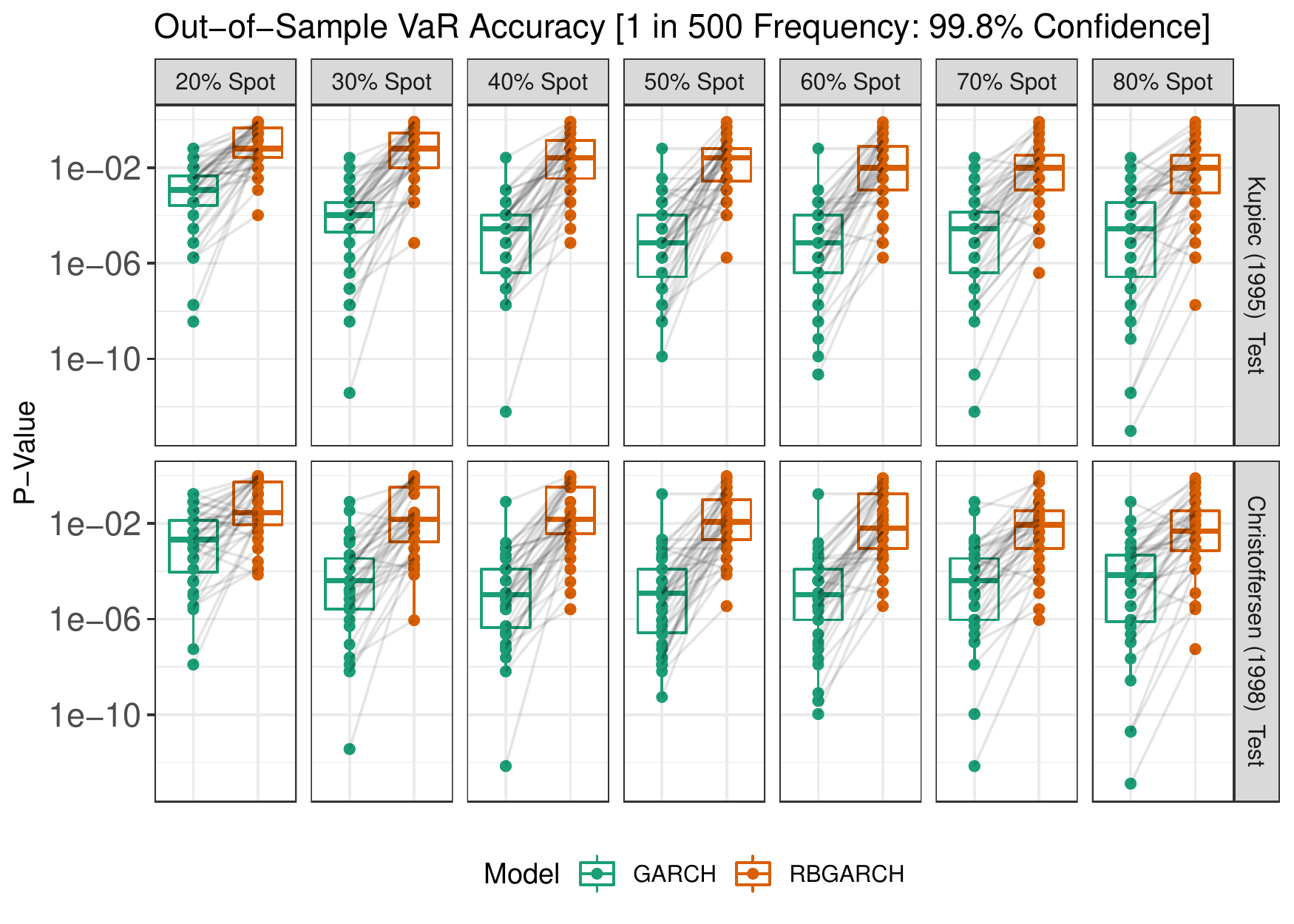}
  \caption{Out-of-sample VaR accuracy measures of the \GARCH and \RBGARCH models at the 99.8\% (1 in 500 days) confidence level. Consistent with Figure \ref{app:fig:out_500}, the \RBGARCH model consistently outperforms the \GARCH model, as can be seen from the upward sloping lines connecting corresponding $p$-values from the two models.}
  \label{app:fig:out_var_paired}
\end{figure}

\begin{figure}[b]
  \centering
  \includegraphics{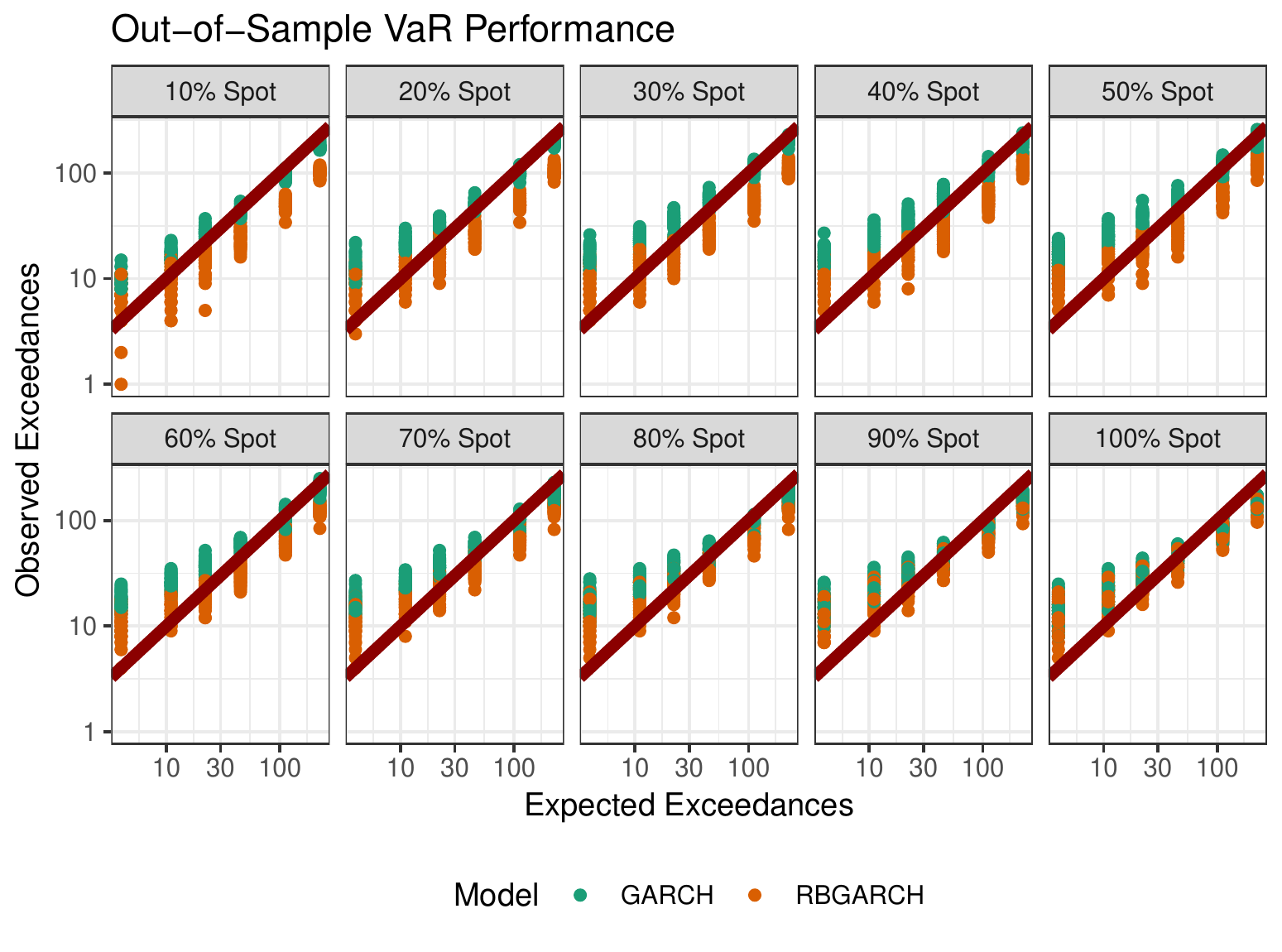}
  \caption{Assessment of the Out-of-Sample VaR estimates of the \GARCH and \RBGARCH models for a range of sample portfolios. The \GARCH model typically underestimates the true VaR (lies above the red line) while the \RBGARCH model is more accurate, but with a conservative bias (lies below the red line).}
  \label{app:fig:out_sample_var}
\end{figure}

Figures \ref{app:fig:out_500}, \ref{app:fig:out_200}, \ref{app:fig:out_var_paired}, \ref{app:fig:out_sample_var} repeat the analysis of the previous subsection on the out-of-sample VaR forecasts. As described in Section \ref{sec:application}, these forecasts are obtained by filtering the volatility for each posterior sample and then calculating an overall VaR estimate by marginalizing over the posterior samples. Not surprisingly, our out-of-sample forecasts are generally less accurate than our in-sample estimates. Despite this, we see essentially the same results as before: the \RBGARCH model consistently outperforms the \GARCH model, doing particularly well on portfolios with large amounts of both spot and futures. When the \RBGARCH is mis-calibrated, it tends to be slightly conservative, but overall it is quite accurate. 

The systematic underestimation of standard GARCH models is well-known to practitioners and model based estimates of VaR are typically multiplied by a ``fudge factor'' to account for this. In 1996, the Basel Committee on Banking Supervision recommended the use of a fudge factor of 3 \citep[Paragraph B.4(j)]{BIS:1996}, cementing the widespread use of this sort of \emph{ad hoc} adjustment. While this style of adjustment is sufficient for conservative risk management, if applied uncritically it can greatly increase the financial burden imposed on market participants due to the increased capital requirements associated with higher VaR levels. The superior performance of the \RBGARCH model suggests that a fudge factor may not be necessary if additional information is used to improve out-of-sample predictive performance.

\clearpage

\section{Computational Details} \label{app:computation}
In this section, we give additional details of the computational approach used to estimate Model \eqref{eqn:model}. To perform estimation, we use the No-U-Turn Sampler variant of Hamiltonian Monte Carlo \citep{Hoffman:2014,Neal:2011,Betancourt:2017a} as implemented in \texttt{Stan} \citep{Carpenter:2017}. Hamiltonian Monte Carlo is particularly well-suited for models such as ours, as it can take advantage of gradient information to more efficiently explore high-dimensional and highly-correlated posterior distributions. The relevant \texttt{Stan} code is given in Section \ref{app:code}.  Section \ref{app:mcmc} describes \emph{ex post} diagnostics to assess the efficiency of Hamiltonian Monte Carlo for our analysis. We confirm that \texttt{Stan} can sample effectively from the posterior using the Simulation-Based Calibration (SBC) approach of \citet{Talts:2018}, as described in Section \ref{app:sbc}. Finally, we examine the finite sample estimation performance of our model in Section \ref{app:power}.

\subsection{\texttt{Stan} Code} \label{app:code}
The probabilistic programming language \texttt{Stan} \citep{Carpenter:2017} provides a modeling language for high-performance Markov Chain Monte Carlo (`MCMC'), using the No-U-Turn Sampler variant of Hamiltonian Monte Carlo \citep{Hoffman:2014}. Unless otherwise stated, \texttt{Stan} was used for all posterior inference reported in this paper. The \texttt{Stan} manual \citep{Stan:2017} provides an exhaustive introduction to the use of \texttt{Stan}. The \texttt{Stan} code presented below was used to fit Model \eqref{eqn:model}. The \texttt{Stan} code for the univariate model used for comparisons in Section \ref{sec:application} can be obtained by removing certain sections of this code. 

\texttt{Stan} does not provide a built-in \emph{multivariate} skew-normal density, so we implement it as a user-defined function. For computational reasons, we model the marginal variances and the correlation matrix separately, so we use a four-parameter formulation of the multivariate skew-normal density. Our specification is essentially that of \citet{Azzalini:1999}, except we use the Cholesky factor of the correlation matrix, which we denote $\Omega$, instead of the full correlation matrix (which they denote $\Omega_z$) to more efficiently evaluate the multivariate normal density. Additionally, we denote the marginal variances by $\sigma$, as opposed to $\omega$.
\begin{minted}[mathescape, fontsize=\footnotesize]{stan}
  functions{
    real multi_skew_normal_lpdf(vector y, vector mu, vector sigma, vector alpha, matrix omega){
        real retval = 0;
        int K = rows(y);

        retval += multi_normal_cholesky_lpdf(y | mu, diag_pre_multiply(sigma, omega));

        for (i in 1:K){
            retval += normal_lcdf(dot_product(alpha, (y - mu) ./ sigma) | 0, 1);
        }
        return retval;
    }
}
\end{minted}
We combine the market (Henry Hub futures) and asset (non-Henry spot) returns into a $T$-length array of 2-vectors.
\begin{minted}[mathescape, fontsize=\footnotesize]{stan}
data {
    int<lower=1> T;
    vector[T] return_market;
    vector[T] return_asset;
    vector<lower=0>[T] realized_vol_market;
}
transformed data{
    vector[2] returns[T];

    for(t in 1:T){
        returns[t][1] = return_market[t];
        returns[t][2] = return_asset[t];
    }
}
\end{minted}
As discussed above, we use the Cholesky factor of the correlation matrix for computational efficiency. This allows for more efficient evaluation of the multivariate normal density (by avoiding an expensive determinant calculation) and allows \texttt{Stan} to use a positive (semi)-definiteness-enforcing transformation automatically \citep{Pinheiro:1996}.
\begin{minted}[mathescape, fontsize=\footnotesize]{stan}
parameters {
    vector[2] mu;
    vector[2] alpha;
    cholesky_factor_corr[2] L;
\end{minted}
We enforce stationarity by constraining the parameters to sections of the parameter space which yield a stationary process. Note that we do not use variance targeting.
\begin{minted}[mathescape, fontsize=\footnotesize]{stan}
    real<lower=0> omega_market;
    real<lower=0,upper=1> gamma_market;
    real<lower=0,upper=(1-gamma_market)> tau_1_market;
    real<lower=0,upper=(1-gamma_market-tau_1_market)> tau_2_market;
\end{minted}
We treat the initial volatility as an unknown parameter to be inferred. In practice, this is not particularly important since we use a long (250 day) window to fit our model.
\begin{minted}[mathescape, fontsize=\footnotesize]{stan}
    real<lower=0> sigma_market1;

    real<lower=0> omega_asset;
    real<lower=0,upper=1> beta_asset;
    real<lower=0,upper=(1-beta_asset)> gamma_asset;
    real<lower=0,upper=(1-beta_asset-gamma_asset)> tau_1_asset;
    real<lower=0,upper=(1-beta_asset-gamma_asset-tau_1_asset)> tau_2_asset;

    real<lower=0> sigma_asset1;

    real xi;
    real phi;
    real<lower=0> delta_1_rv;
    real<lower=0> delta_2_rv;
    real<lower=0> rv_sd;
}
\end{minted}
Because the instantaneous volatilities in a GARCH model are deterministic, conditional on the returns, model parameters, and initial condition, we calculate them in the \texttt{transformed parameters} block. Note the use of local variables \texttt{sigma\_market} and \texttt{sigma\_asset} to reduce memory pressure.
\begin{minted}[mathescape, fontsize=\footnotesize]{stan}
transformed parameters {
    vector<lower=0>[2] sigma[T];
    vector[T] rv_market_mean;

    {
        vector[T] sigma_market;
        vector[T] sigma_asset;

        sigma_market[1] = sigma_market1;
        sigma_asset[1]  = sigma_asset1;

        for(t in 2:T){
            sigma_market[t] = sqrt(omega_market +
                                   gamma_market * square(sigma_market[t - 1]) +
                                   tau_1_market * fabs(return_market[t-1]) +
                                   tau_2_market * square(return_market[t-1]));

            sigma_asset[t] = sqrt(omega_asset +
                                  gamma_asset * square(sigma_asset[t - 1]) +
                                  beta_asset  * square(sigma_market[t - 1]) +
                                  tau_1_asset * fabs(return_asset[t-1]) +
                                  tau_2_asset * square(return_asset[t-1]));
        }

        for(t in 1:T){
            sigma[t][1] = sigma_market[t];
            sigma[t][2] = sigma_asset[t];
        }

        rv_market_mean = xi + phi * sigma_market +
                              delta_1_rv * fabs(return_market) +
                              delta_2_rv * square(return_market);
    }
}
\end{minted}
The priors from Table \ref{tab:priors} are used. Note that \texttt{Stan} uses a mean-standard deviation parameterization of the normal distribution as opposed to the more common mean-variance or mean-precision parameterizations.
\begin{minted}[mathescape, fontsize=\footnotesize]{stan}
model {
    // Priors
    mu           ~ normal(0, 1);
    alpha        ~ normal(0, 1);
    L            ~ lkj_corr_cholesky(1);
    beta_asset   ~ normal(0, 1);

    // Market (Henry Hub) GARCH Dynamics
    // $\sigma_{t}^2 = \omega + \gamma \sigma_{t-1}^2 + \tau_1 |r_{t-1}| + \tau_2 r_{t-1}^2$
    omega_market ~ normal(0.002, 0.025);
    gamma_market ~ normal(0.8, 0.6);
    tau_1_market ~ normal(0, 0.1);
    tau_2_market ~ normal(0.1, 0.7);

    // Asset (non-Henry Hub Spot) GARCH Dynamics
    // $\sigma_{t}^2 = \omega + \gamma \sigma_{t-1}^2 + \tau_1 |r_{t-1}| + \tau_2 r_{t-1}^2$
    omega_asset ~ normal(0.002, 0.025);
    gamma_asset ~ normal(0.8, 0.6);
    tau_1_asset ~ normal(0, 0.1);
    tau_2_asset ~ normal(0.1, 0.7);

    // Realized volatility dynamics
    // $\varsigma_t = \xi + \phi \sigma_t + \delta_1 |r_t| + \delta_2 r_t^2 + \mathcal{N}(0, \nu^2)$ 
    xi  ~ normal(0.02, 0.6);
    phi ~ normal(15, 60);
    delta_1_rv ~ normal(1.15, 8);
    delta_2_rv ~ normal(1.15, 14);
    rv_sd ~ normal(0.05, 0.25);
\end{minted}
A weak data-dependent prior is on the initial conditions of the GARCH volatility.
\begin{minted}[mathescape, fontsize=\footnotesize]{stan}
    // Initialize initial vol with weak prior
    sigma_market1 ~ normal(sd(return_asset), 5 * sd(return_asset));
    sigma_asset1  ~ normal(sd(return_asset), 5 * sd(return_asset));

    // Likelihood
    realized_vol_market ~ normal(rv_market_mean, rv_sd);
    for(t in 1:T){
        returns[t] ~ multi_skew_normal(mu, sigma[t], alpha, L);
    }
}
\end{minted}
$\Sigma_t = \text{diag}(\sigma_t) L^TL \text{diag}(\sigma_t)$ is the covariance matrix of the returns model.
\begin{minted}[mathescape, fontsize=\footnotesize]{stan}
generated quantities{
    cov_matrix[2] Sigma[T];
    for(t in 1:T){
        Sigma[t] = tcrossprod(diag_pre_multiply(sigma[t], L));
    }
}
\end{minted}
\subsection{MCMC Diagnostics} \label{app:mcmc}
We propose the use of the Hamiltonian variant of Markov Chain Monte Carlo to estimate Model \eqref{eqn:model}. While Hamiltonian Monte Caro (HMC), like all commonly-used MCMC methods, is guaranteed to \emph{asymptotically} recover the true posterior, one should always carefully assess its performance in a given simulation before performing inference based on its output. \citet{Gelman:2011} give a succinct review of standard MCMC diagnostics. We supplement these with various recently-developed HMC-specific diagnostics, such as the Expected Bayesian Fraction of Missing Information (E-BFMI) \citep{Betancourt:2016} and divergent transitions \citep[Section 6.2]{Betancourt:2017a}. While the use of Hamiltonian Monte Carlo for Realized GARCH models has not been previously examined, our results suggest that it is an efficient and robust sampling scheme for this type of model.

For each run of HMC, we run four separate chains of 2000 iterations.\footnote{Following \texttt{Stan}'s default settings, the first 1000 iterations were used as an adaptation period and discarded, while the second 1000 iterations were stored. No thinning was performed.} While 2000 iterations may seem insufficient, particularly as compared to the tens of thousands typically used with other samplers, HMC typically mixes far more efficiently than other methods, reducing the total number of iterations required. As we will see below, this is sufficient for our problem. To increase the robustness of our results, we used a slightly higher target adaptation rate ($\texttt{adapt\_delta} = 0.99$) and maximum tree-depth ($\texttt{max\_treedepth} = 12$) than \texttt{Stan}'s default settings ($0.8$ and $10$ respectively). On the vast majority of our fits, no divergent transitions were encountered, the maximum treedepth was never hit, and the average E-BMFI across chains was above 0.9 (values below 0.2 are typically considered indicative of a sampling pathology). Taken together, these results suggest that the sampler was able to efficiently explore the posterior distribution. 

Having confirmed that the sampler was able to efficiently explore the posterior distribution, we are now in a position to assess whether the sampler output provides enough precision for subsequent analyses. We compute both the effective sample size and the split-$\hat{R}$ diagnostic \citep[Section 11.4-11.5]{Gelman:2013} for each of the parameters of our model as well as the estimated volatilities $\sigma_t$. Our results are presented in Table \ref{tab:mcmc}. The diagnostics suggest that we are able to explore the posterior efficiently and that our results are sufficiently precise for downstream analyses. 

\subsection{Simulation-Based Calibration}\label{app:sbc}
\begin{figure}
\centering
\includegraphics[width=\textwidth]{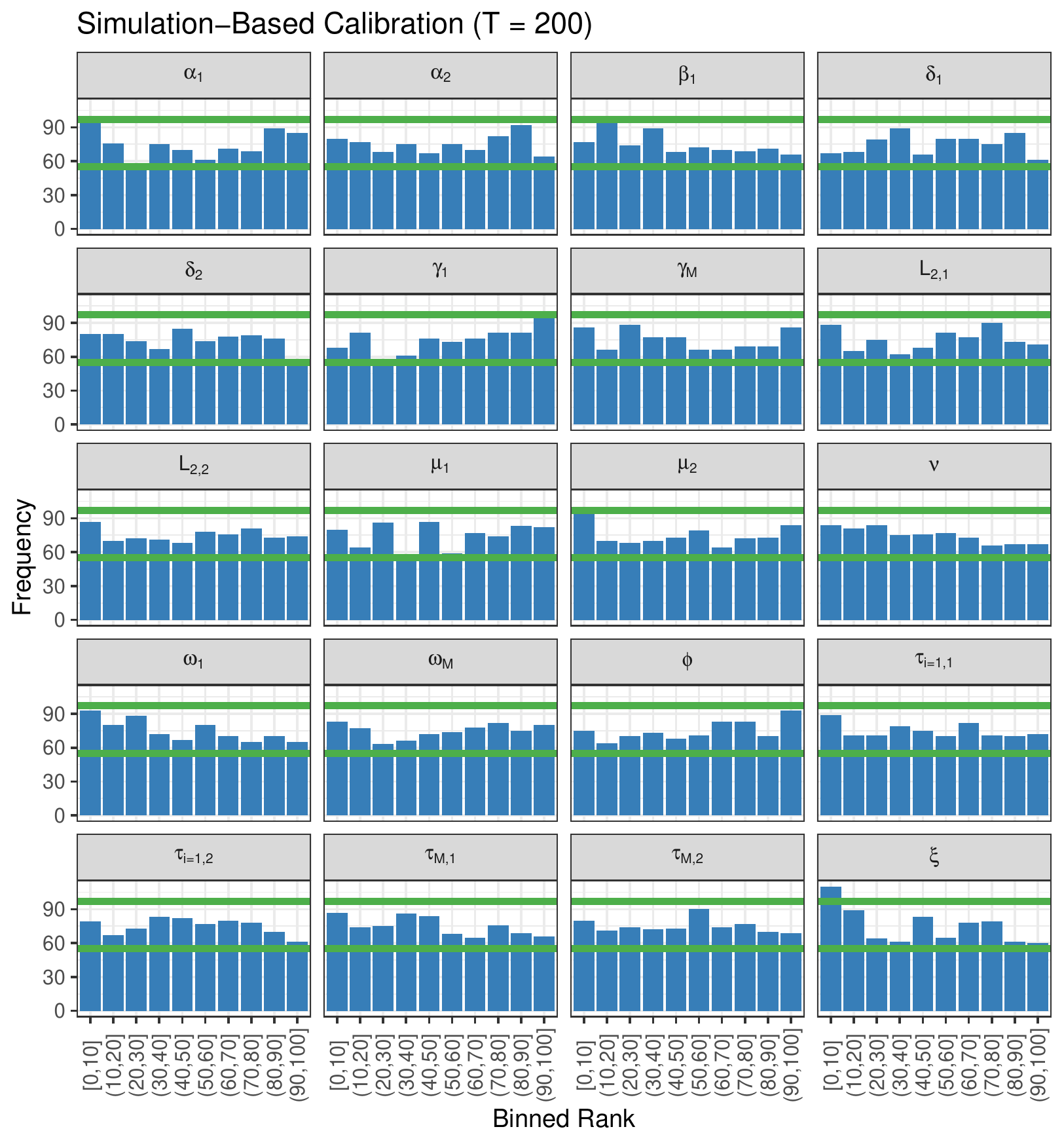}

\caption{Simulation-Based Calibration Analysis of Hamiltonian Monte Carlo applied to Model \eqref{eqn:model} with $T = 200$. The blue bars indicate the number of SBC samples in each rank bin while the green bars give a 99\% marginal confidence band for the number of samples in each bin. While there is significant variability in the histograms, commensurate with a complex model, we do not observe any \emph{systematic} deviations.}
\label{fig:sims_sbc}
\end{figure}

\citet{Talts:2018} propose a general scheme for validating Bayesian inference, based on the idea of the \emph{data-averaged posterior}. They note that, if the parameters are indeed drawn from the prior, the average of the \emph{exact} posterior distributions is exactly the prior. As such, any deviation between the data-averaged posterior produced by a sampling scheme and the prior indicates biases in the sampling scheme. Applying this technique to Model \eqref{eqn:model} yields the results in Figure \ref{fig:sims_sbc}, which show no systematic deviations from uniformity, suggesting that our inference is unbiased and that our sampling scheme is well-suited for the model. 

\subsection{Additional Simulation Studies} \label{app:power}
In this section, we provide additional evidence that Model \eqref{eqn:model} is able to effectively capture the underlying GARCH dynamics, even with relatively small data sizes. Since we are concerned with performance across a wide range of possible parameter values, we use wider priors for these simulation studies than we do in the main body of this paper. In particular, we use a weakly informative $\mathcal{N}(0, 1)$-prior on all GARCH parameters for each study in this section.

\begin{figure}
\begin{center}
\includegraphics[width=\textwidth]{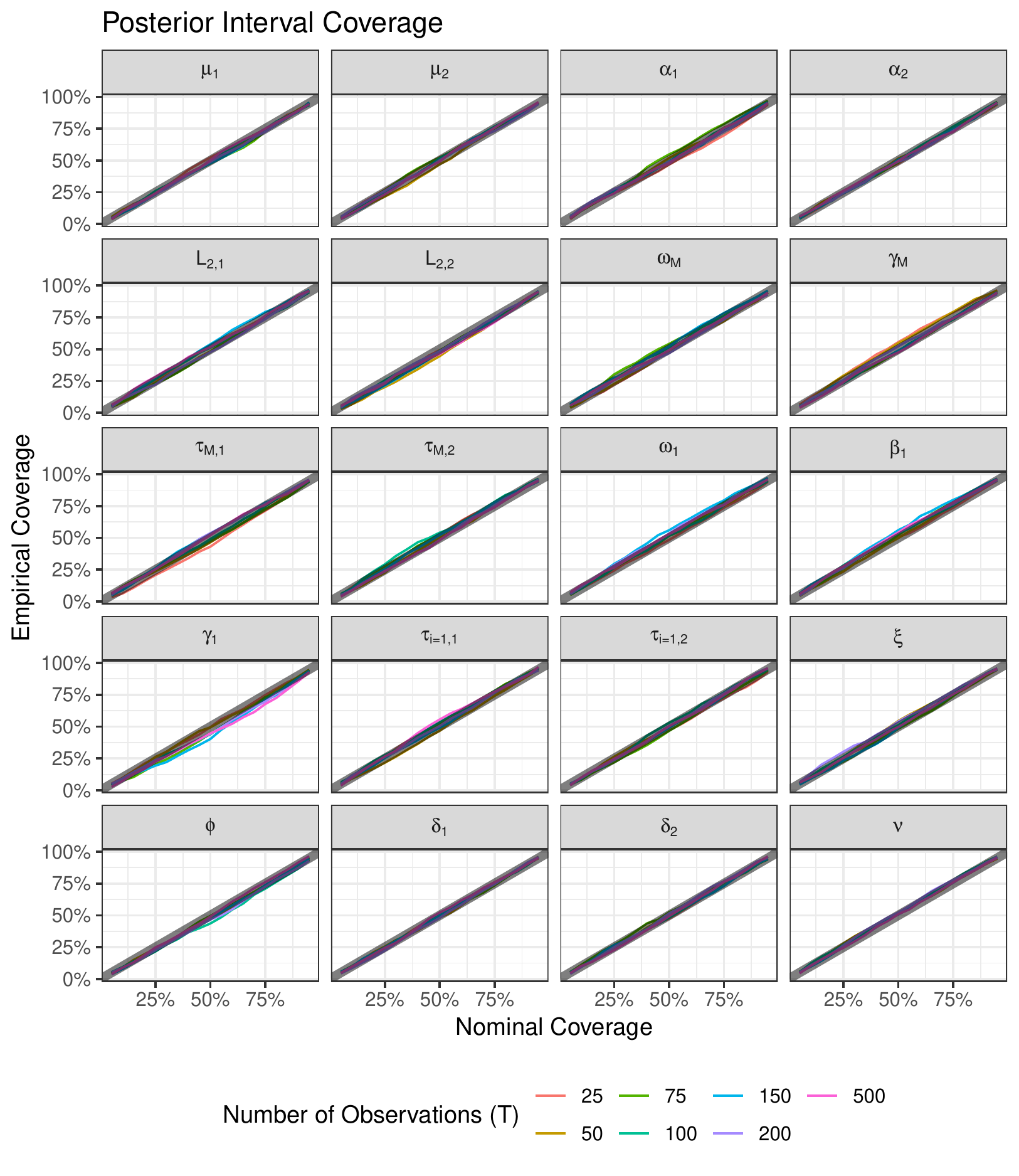}
\end{center}
\caption{Empirical coverage probabilities of posterior credible intervals associated with Model \eqref{eqn:model} under a weakly informative prior distribution. Because we drew the parameters from the prior, these intervals are correctly calibrated.}
\label{fig:sims_coverage}
\end{figure}

\begin{figure}
\begin{center}
\includegraphics[width=\textwidth]{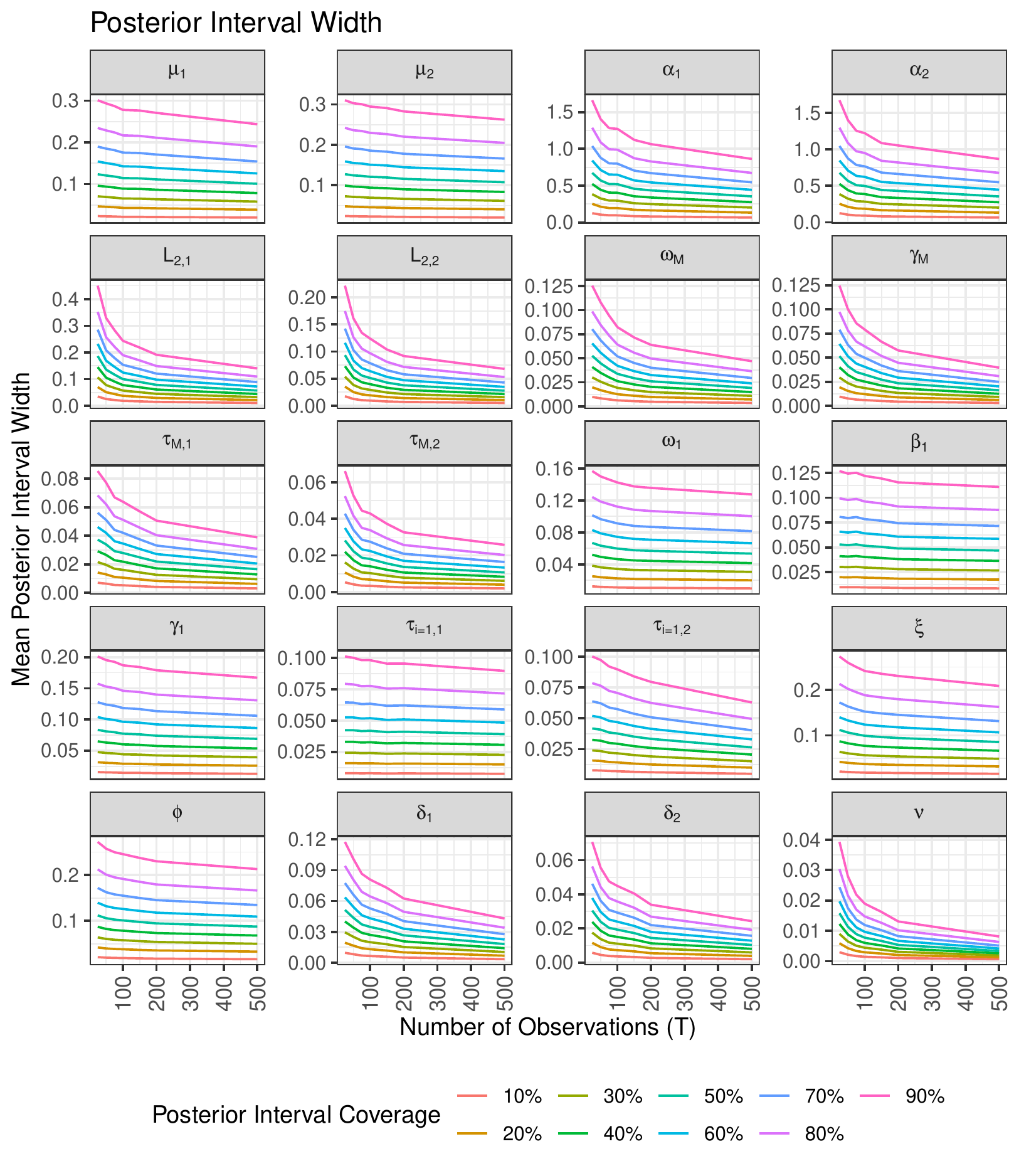}
\end{center}
\caption{Mean width of symmetric posterior intervals associated with Model \eqref{eqn:model} under a weakly informative prior distribution. We observe standard $n^{-1/2}$-type convergence for all parameters as the length of the sample period ($T$) increases. Somewhat surprisingly, we note that parameters associated with the mean model ($\mu, \alpha$) are typically more uncertain than the parameters of the GARCH dynamics.}
\label{fig:sims_width}
\end{figure}

The natural first question to ask of any Bayesian model is whether the posterior credible intervals are indeed accurately computed. If the parameters are drawn from the prior distribution, the posterior credible intervals should be perfectly calibrated and, \emph{ceteris paribus}, should decrease in length as the sample size increases. Figures \ref{fig:sims_coverage} and \ref{fig:sims_width} confirm that our posterior credible intervals are correctly calculated and well calibrated. \citet{Hansen:2012} demonstrated that the univariate Realized GARCH framework satisfies the conditions for standard ($1/\sqrt{T}$) asymptotic convergence of the Quasi-Maximum Likelihood Estimate. Figure \ref{fig:sims_width} suggests that Model \eqref{eqn:model} is similarly well-behaved and that Bernstein-von Mises-type convergence rates can be obtained.


A close examination of Figure \ref{fig:sims_width} will reveal that parameters associated with the mean model -- namely the $\mu$ and $\alpha$ parameters -- have wider posterior credible intervals than the parameters of the GARCH and realized volatility processes ($\omega$, $\gamma$, $\tau$, $\beta$, \emph{etc.}) While this may seem counter-intuitive at first, it appears to be a consequence of the relatively high-levels of volatility considered, which make precise inference on the mean difficult. Despite this difficulty, our simulations suggest that Model \ref{eqn:model} is able to recover the underlying dynamics effectively even with relatively small samples.

\begin{table}[phtb]
\centering
\footnotesize
\begin{tabular}{crr}
\toprule
{\bf Parameter} & {\bf Effective Sample Size ($\widehat{n_{\text{eff}}}$)} & {\bf Potential Scale Reduction Factor ($\widehat{R}$)}  \\ 
\midrule
$L_{21} = \textsf{chol}(\Omega)_{21}$ & 2704.386 & 0.999 \\
$L_{22} = \textsf{chol}(\Omega)_{22}$ & 2705.967 & 0.999 \\
$\omega_M$ & 1491.062 & 1.001 \\
$\gamma_M$ & 2345.653 & 1.000 \\
$\tau_{1, M}$ & 2223.725 & 1.000 \\
$\tau_{2, M}$ & 2705.410 & 1.001 \\
$\sigma_M$ & 2051.442 & 1.000\\
\midrule
$\omega_i$ & 1994.433 & 1.000 \\
$\beta_i$ & 2604.040 & 1.000\\
$\gamma_i$ & 2625.324 & 1.000 \\
$\tau_{1, i}$ & 2331.136 & 1.000\\
$\tau_{2, i}$ & 2372.701 & 1.000\\
$\sigma_i$ & 2084.207 & 1.000\\
\midrule
$\xi$ & 1715.700 & 1.001\\
$\phi$ & 1662.953 & 1.001\\
$\delta_1$ & 2395.295 & 1.000\\
$\delta_2$ & 2297.632 & 1.000 \\
$\nu$ & 4000.000 & 1.000\\
\bottomrule
\end{tabular}
\caption{Markov Chain Monte Carlo Diagnostics for Hamiltonian Monte Carlo estimation of Model \ref{eqn:model}. These diagnostics are taken from a representative run of four chains of 2000 iterations each. Reported values for $\sigma_M$ and $\sigma_i$ are averages over the entire fitting period.}
\label{tab:mcmc}
\end{table}

\section{Additional Background} \label{app:additional_background}

In this section, we give some additional background for the reader who may be interested in learning more about volatility models or the structure of LNG markets.

\subsection{Volatility Models}

The first latent volatility process model was the Autoregressive Conditional Heteroscdasticity (ARCH) model of \citet{Engle:1982}, which modeled the instantaneous volatility $\sigma_t^2$ as a weighted sum of the $q$-previous (standardized) returns. \citet{Bollerslev:1986} proposed the \emph{Generalized} ARCH (GARCH) model which instead uses an autoregressive moving average (ARMA) model for the instantaneous volatility, combining previous volatility levels and (standarized) returns. As demonstrated by \citet{Bollerslev:1986}, the use of previous volatility levels significantly reduces the amount of history required, giving a more accurate and more parsimonious model. For this reason, GARCH-type models have generally supplanted ARCH models in applied work.

GARCH-type models have had an enormous impact on financial econometrics and many variants have been proposed, including the integrated GARCH (I-GARCH) of \citet{Engle:1986}; the exponential GARCH (E-GARCH) of \citet{Nelson:1991}; the GJR-GARCH of \citet{Glosten:1993}, which allows for asymmetric positive and negative effects; and the asymmetric power GARCH of \citet{Ding:1993}, which introduces a Box-Cox transform \citep{Box:1964} into the model specification and unifies several previous proposals. The family GARCH model of \citet{Hentschel:1995} gives a very general specification comprising a wide range of GARCH variants. More recent variants attempt to decompose latent volatilities into long- and short-term components, yielding the additive and multiplicative component GARCH models of \citet{Engle:1999} and \citet{Engle:2012}, respectively. The glossary of \citet{Bollerslev:2010} provides a useful and comprehensive list of variants. 

Multivariate extensions of GARCH models are equally numerous and we refer the reader to the survey of \citet{Bauwens:2006} for a more detailed review. The simplest multivariate GARCH model is perhaps that of \citet{Bollerslev:1990}, who assumes a constant conditional correlation (CCC) matrix among the various assets. \citet{Engle:2002} extended the CCC model was extended to allow for slowly-varying conditional correlations, yielding the popular dynamic conditional correlation (DCC) specification. Our proposed model uses the CCC approach, though the possibility of misspecification is mitigated by our rolling refitting strategy which yields an approximate DCC specification.

A popular alternative to GARCH models is the class of ``stochastic volatility'' models first introduced by \citet{Kim:1998}. These models allow the volatility to evolve according to a (typically independent) stochastic process, giving their name. This is in contrast to GARCH-type models where the next day's volatility is deterministic, conditional on the current return and volatility history. We do not review these models in detail here, instead referring the reader to the paper of \citet{Asai:2006} and the discussion in \citet{Vankov:2019}

A commonly noted shortcoming of both SV- and GARCH-type models is their slow responsiveness to rapid changes in the volatility level \citep{Engle:2002b,Anderson:2003,Anderson:2005}. This slowness is a consequence of the ``smoothing'' nature of both SV and GARCH models, which balance the information provided by a single day's return against a long history. By using several volatility measures, it becomes possible to weight the current time period more heavily and to develop more responsive models, as shown by, \emph{e.g.}, \citet{Visser:2011} and \citet{Anderson:2011} for GARCH models and \citet{Takahashi:2009} for SV models. 

Driven by market conventions, the most common additional volatility measures are those based on OHLC (Open, High, Low, Close) data, such as the high-low range proposed by \citet{Parkinson:1980}, the open-close difference proposed by \citet{Garman:1980}, or high, low, and close data as suggested by \citet{Rogers:1991}. The estimator of \citet{Yang:2000} is optimal (minimum variance unbiased) among estimators based solely on OHLC data. If higher frequency data is available, even more accurate estimators have been proposed, though their efficacy is very sensitive to the structure of the market being considered \citep{Barndorff-Neilsen:2002,Barndorff-Neilsen:2003,Zhang:2005,Barndorff-Nielsen:2011}.

While realized volatility measures can be used to improve GARCH estimation by replacing the standardized squared return with an improved estimate, this naive approach does not capture the structural relationships among different volatility measures. Several classes of ``complete'' models which jointly model prices and realized volatilty measures have been proposed, including the  Multiplicative Error Model (MEM) framework of \citet{Engle:2006} and the High-Frequency Based Volatility (HEAVY) framework of  \citet{Shephard:2010}. In this paper, we consider the Realized GARCH framework of \citet{Hansen:2012,Hansen:2014,Hansen:2016}, which is among the most flexible specification proposed to date. In addition to standard volatility estimation, the realized GARCH framework has proven useful for risk management, with \citet{Watanabe:2012} demonstrating its usefulness in conditional quantile estimation and \citet{Contino:2017,Wang:2018} showing its usefulness for estimating VaR and Expected Shortfall \citep{Artzner:1999,Acerbi:2002,Jorion:2006}.

\subsection{Natural Gas Markets}

The past 20 years have seen significant developments in the structure and size of US LNG markets. Much of this development has been driven by the development of shale gas extraction technologies, in particular hydraulic fracturing (``fracking''), which have lowered production costs and increased demand for LNG and LNG derivatives \citep{Caporin:2017}. At the same time, increasing awareness of the environmental impacts of various energy sources has prompted additional investment into LNG: while a fossil fuel and not renewable, LNG is widely considered to be a cleaner energy source than coal and a possible bridge to a fully renewable energy sector. Over the same period, investor interest in LNG and other commodities has increased, spurring the growth of liquid futures and derivatives markets \citep{Tang:2012}. 

\citet{Li:2019}, following \citet{Narayan:2015}, identifies September 2008 as the ``coming of age'' of modern LNG markets. Modern LNG markets are characterized by active, but relatively stable, spot trading at a large number of hubs and by informative and liquid futures markets. For historical reasons, Henry Hub plays a particularly prominent role in LNG markets, \emph{inter alia} serving as the reference price for much of the LNG derivatives market, it is far from the only hub of economic interest. Practitioners recognize over one hundred trading hubs in the continental United States and Canada, with the Chicago Citygate, Algonquin Citygate (serving the Boston area), Opal (Lincoln County, Wyoming), Southern California, and NOVA (Alberta, Canada) hubs being particularly closely monitored by market participants. (Strictly speaking, some of these are commonly reported regional average indices, not physical hubs, but we will continue to refer to them as hubs.)

\citet{Mohammadi:2011} gives a detailed survey of the pricing structure from well-heads (extraction) to end-consumers, while \citet{Hou:2018} and \citet{Hailemariam:2019} attempt to quantify the effect of supply and demand shocks on observed prices. Unlike equities, LNG spot and futures prices do not move in perfect synchrony due to storage and transpot costs. The relationship between LNG spot and futures prices was investigated by \citet{Ghoddusi:2016} who found, \emph{inter alia} that short-maturity futures are Granger-causal for physical prices, a fact which is consistent with our findings on the usefulness of futures realized volatility in spot price volatility modeling. 
measurable impact on observed price dynamics.

The relationship between the prices of LNG and other commodities has been particularly well-studied, and has been found to be surprisingly complex: \citet{Hartley:2008} found important dependencies on foreign exchange rates, spot inventories, and unpredictable (weather) shocks; \citet{Brigida:2014} and \citet{Batten:2017} found that technological advancements weakened, but did not eliminate the underlying relationship; and \citet{Caporin:2017} investigated the effect of the shale revolution on the relationship. 

\clearpage
\printbibliography[title={Additional References}]
\end{refsection}
\end{document}